\documentclass[A4paper,useAMS,usenatbib]{mnras}

\usepackage[T1]{fontenc}
\usepackage{ae,aecompl}
\usepackage{graphicx}
\usepackage{rotating}
\usepackage{float}
\usepackage{array}
\usepackage{mathenv}
\usepackage{multirow}
\usepackage{url}
\usepackage{subcaption}
\usepackage{comment}
\usepackage{mathptmx}
\defcitealias{Mingo2014}{Paper I}
\newcommand{\edit}{}

\begin{document}

  \title[An X-ray survey of the 2Jy sample II]{An X-ray survey of the 2Jy sample. II: X-ray emission from extended structures}
   \author[B. Mingo et. al.]{B. Mingo$^{1}$\thanks{E-mail:bmingo@extragalactic.info}, M. J. Hardcastle$^{2}$, J. Ineson$^{3}$, V. Mahatma$^{2}$, J. H. Croston$^{4}$, D. Dicken$^{5}$, \newauthor D. A. Evans$^{6}$, R. Morganti$^{7,8}$, and C. Tadhunter$^{9}$ \\
   		$^{1}$Department of Physics and Astronomy, University of Leicester, University Road, Leicester LE1 7RH, UK\\
   		$^{2}$Centre for Astrophysics Research, School of Physics, Astronomy \& Mathematics, University of Hertfordshire, College Lane, Hatfield AL10 9AB, UK\\
		$^{3}$School of Physics and Astronomy, University of Southampton, Southampton SO17 1SJ, UK\\
		$^{4}$School of Physical Sciences, The Open University, Walton Hall, Milton Keynes MK7 6AA, UK\\
		$^{5}$CEA-Saclay, F-91191 Gif-sur-Yvette, France\\
		$^{6}$Harvard-Smithsonian Center for Astrophysics, 60 Garden Street, Cambridge, MA 02138, USA\\
		$^{7}$ASTRON, the Netherlands Institute for Radio Astronomy, Postbus 2, 7990 AA, Dwingeloo, The Netherlands\\
		$^{8}$Kapteyn Astronomical Institute, University of Groningen, P.O. Box 800, 9700 AV Groningen, The Netherlands\\
		$^{9}$Department of Physics and Astronomy, University of Sheffield, Hounsfield Road, Sheffield S3 7RH, UK }
   \date{Received ; accepted}

\maketitle
\begin{abstract}
The 2Jy sample is a survey of radio galaxies with flux densities above 2 Jy at 2.7 GHz. As part of our ongoing work on the southern subset of 2Jy sources, in paper I of this series we analysed the X-ray cores of the complete 2Jy sample with redshifts $0.05<z<0.7$. For this work we focus on the X-ray emission associated with the extended structures (jets, lobes, and environments) of the complete subset of 2Jy sources with $0.05<z<0.2$, that we have observed with \textit{Chandra}. We find that hotspots and jet knots are ubiquitous in FRII sources, which also inhabit systematically poorer environments than the FRI sources in our sample. Spectral fits of the hotspots with good X-ray statistics invariably show properties consistent with synchrotron emission, and we show that inverse-Compton mechanisms under-predict the X-ray emission we observe by 1--2 orders of magnitude. Inverse-Compton emission is detected from many of the lobes in our sample, and we find that the lobes of the FRII sources show magnetic fields lower by up to an order of magnitude than expected from equipartition extrapolations. This is consistent with previous results, which show that most FRII sources have electron energy densities higher than minimum energy requirements.

\end{abstract}

   \begin{keywords}
   		galaxies: active --
		X-rays: galaxies --
		radio continuum: galaxies --
   \end{keywords}

%

\section{Introduction}\label{Introduction}

The 2Jy sample of radio galaxies\footnote{\url{http://2Jy.extragalactic.info/2Jy_home_page.html}}, as defined by \citet{WP1985}, includes all the galaxies with flux greater than 2Jy at 2.7 GHz. Over the last twenty years we have obtained and studied in detail uniform data for the complete subset of Southern sources defined by \citet{Tadhunter1993} and \citet{Morganti1993} ($\delta <+10^{\circ}$) and, especially, the steep-spectrum \edit{($\alpha>0.5$, where $\alpha$ is the radio spectral index, such that $S_{\nu} \propto \nu^{-\alpha}$)} subsample defined by \citet{Dicken2008}, which contains 47 objects and is statistically complete for redshifts $0.05<z<0.7$ \citep[see][]{Morganti1993,Morganti1999,Tadhunter1993,Tadhunter1998,Inskip2010,RamosAlmeida2011b,Dicken2008,Dicken2009,Dicken2012,Dicken2014,Mingo2014}. Most recently, in the first paper of this series \citep[][hereafter, Paper I]{Mingo2014}, we analysed the X-ray cores of the 2Jy sources in the subset of \citet{Dicken2008}, using data from \textit{Chandra} and \textit{XMM-Newton}, and found our results to be in good agreement with those of \citet{Hardcastle2006,Hardcastle2009} on the 3CRR radio galaxies.

In this work we focus on the extended X-ray emission (jets, hotspots, and lobes) and the environments of the $0.05<z<0.2$ subset of sources that we have observed with \textit{Chandra}, whose nuclei we studied in paper I. Our knowledge of X-ray jets \citep[see e.g. the review by][]{Worrall2009} and hotspots \citep[e.g.][]{Hardcastle2004,Hardcastle2007,Massaro2010,Massaro2015} has certainly improved over the last two decades, as has our understanding of the environment in which radio galaxies live \citep[e.g.][]{Belsole2007,Croston2008b,Ineson2013,Ineson2015}, but the samples of radio galaxies with available detailed observations are still relatively small, and more work needs to be done to understand their extended structures and how they co-evolve with the hosts \citep[see also the recent review by][]{Tadhunter2016}. The 2Jy sample is important in that it is not only statistically complete, but uniformly observed, with long \textit{Chandra} and \textit{XMM-Newton} exposures ($\sim20$ kiloseconds on average) that allow a detailed spectroscopic study of some of the most important structures.

The traditional radio classification, defined by \citet{FR1974}, divides the sources according to their radio structure, into centre-brightened (FRI), and edge-brightened (FRII) classes. This division is tied to the total radio luminosity of the source, with FRIs being less luminous and FRIIs being more so; Fanaroff \& Riley's transition corresponds to a power of $10^{25}$ W Hz$^{-1}$ at 1.4 GHz. The radio luminosity in turn is expected to be related to the intrinsic jet power $Q$, but radio luminosity must also be affected by other factors, including a source's age and the density of its environment \citep[e.g.][]{Hardcastle2013,Hardcastle2014,English2016}, so that morphology and radio luminosity are not always reliable estimators of intrinsic jet power. FRI jets are known to decelerate from relativistic to non-relativistic speeds on kpc scales \citep[e.g.][]{LaingBridle2014}, which implies relatively substantial entrainment of external material. In general terms, the standard explanation for the FRI/FRII dichotomy \citep[e.g.][]{Bicknell1995} is that FRI jets, which are less intrinsically powerful (lower $Q$), are decelerated by entrainment, to transonic speeds before leaving the environment of the host galaxy, while FRII jets are powerful enough to retain supersonic (relativistic) speeds on scales of tens of kpc. The FRI/FRII division would thus be a function of both environment and intrinsic jet power \edit{\citep[but see also e.g.][]{Tchekhovskoy2016}}.

Consistent with this, FRI sources in flux-limited samples have long been thought to inhabit relatively dense environments \citep[e.g.][]{PrestagePeacock1988}, although this seems to change for low-luminosity FRI LERGs \citep[low excitation radio galaxies, see][]{Ineson2015}, and there is evidence from pressure balance arguments \citep[e.g.][]{Croston2007,Mingo2012} that their lobes may contain a substantial non-radiating component, and as such depart substantially from an assumption of energy equipartition between the magnetic field and the electrons in the lobes \citep{Croston2008b}. FRIIs inhabit sparser environments \citep[e.g.][]{Ineson2013,Ineson2015}, and their lobes are closer to equipartition (\citealt{Croston2005}, but see also \citealt{Harwood2016}), though they can drive strong shocks into their surroundings as well \citep{Croston2011}. Pressure balance arguments do not require a substantial non-radiating component in the lobes of many FRIIs \citep{Hardcastle2002} and these differences in particle content mean that, a priori, the same correlations between jet kinetic energy and radio luminosity cannot be applied across both populations \citep[although the dependence of radio luminosity on environment compensates for this fact to some extent, see also][]{Hardcastle2013,Hardcastle2014,Godfrey2013,Godfrey2016,English2016,Mingo2016}.

The FRI/FRII dichotomy should not be confused with the well-known accretion mode dichotomy in radio-loud AGN \citep{Hine1979,Laing1994}. Many FRI also have radiatively inefficient \citep{Narayan1995} nuclei \citep[e.g][]{Hardcastle2006,Hardcastle2007b,Hardcastle2009}, but that is not always the case.  Many, but not all, FRIIs have radiatively efficient \citep[``traditional'' AGN, ][]{Shakura1973} nuclei \citep[however, see e.g. the recent review by][]{Tadhunter2016}. The environmental properties of these sources seem to be tied to their accretion mode, rather than their radio morphology \citep{Ineson2015}. We discussed the nature of the AGN in the 2Jy sample in great detail in \citetalias{Mingo2014}, and use our classifications from that paper in this work. 

Since the energy-loss timescales for relativistic electrons are inversely proportional to their energies, synchrotron emission from radio galaxy lobes is generally detected only at radio frequencies, unless there is an on-going source of particle acceleration. The dominant X-ray emission process from the lobes themselves appears to be inverse-Compton scattering of CMB photons \citep{Feigelson1995,Hardcastle2002,Croston2005}. However, in richer environments (often those of FRIs) the X-ray emission is dominated by thermal bremsstrahlung from the undisturbed large-scale environment and/or shocked gas surrounding the radio source \citep[e.g.][]{Croston2007,Mingo2011,Mingo2012}. One of our objectives in the present paper is to carry out a systematic search for lobe-related emission (inverse-Compton) and extended thermal emission around the 2Jy objects.


Hotspots are the termination points of FRII jets, assumed to be the terminal shocks expected at the end of a supersonic jet \citep{Meisenheimer1989}. Hotspots are regions of intense, on-going particle acceleration, and as such they are bright in the radio, but can be detected at shorter wavelengths as well. In X-rays they often display synchrotron or synchrotron self-Compton spectra \citep{Hardcastle2004,Hardcastle2007}, the latter being more frequent in very luminous hotspots. Often the X-ray hotspots are slightly offset from their radio counterparts, hinting at an underlying complexity in the local environment or the magnetic field. In many sources, including several FRIs, we also see secondary bright spots along the jet. It is likely that some of these so-called knots, which we detect beyond the radio, are also the results of shocks, as they must have on-going particle acceleration to produce synchrotron emission in the optical and X-rays, but others seem to present more diffuse structures and no particle acceleration, indicating, rather, points in which the jet kinetic energy is transferred into particles without the jet being significantly disrupted. These diffuse knots can sometimes be faint in the radio but bright in X-rays \citep[see e.g.][and references therein, for examples of different hotspots and knots and their interactions with the environment]{Hardcastle2004,Hardcastle2007,Hardcastle2008,Massaro2010,Massaro2015,Mack2009,Werner2012,Goodger2010,Kharb2012,Orienti2012,Hardcastle2016,Worrall2016}. It is still not clear what makes some hotspots, knots and jet features X-ray synchrotron sources while others are undetected in the X-rays, and the non-uniform nature of the existing large samples \citep{Hardcastle2004} makes it hard to draw conclusions from observations.

In this paper we use our relatively uniform survey of the $z<0.2$ 2Jy sources to assess the incidence of X-ray hotspots in FRII sources and investigate the mechanisms that produce their X-ray emission, compare the environments we find for FRI and FRII with what we know from the literature, and test the predictions for the inverse-Compton emission in FRIIs against the lobes we detect in X-rays. A detailed study of the large-scale environments of the 2Jy sources, which ties in with some of our results, was carried out by \citet{Ineson2015}. A follow-up study by \citet{Ineson2017} provides further details on the energetics of the 2Jy FRII sources, as part of a larger sample of FRIIs.

For this paper we have used a concordance cosmology with $H_{0}=70$ km s$^{-1}$ Mpc$^{-1}$, $\Omega_{m}=0.3$ and $\Omega_{\Lambda}=0.7$, for compatibility with the results we presented in \citetalias{Mingo2014}. 

%

\section{Data}\label{Data}

\begin{table*}\small
\caption[Objects in the 2Jy sample observed with \textit{Chandra}]{\footnotesize{Objects in the 2Jy sample observed with \textit{\textit{Chandra}} (ACIS-S except for PKS 0625$-$53 and PKS 2135$-$14, which were taken with the ACIS-I), also detailing the radio data used to generate the contours for each source (Figs. \ref{2Jy_0034_f} to \ref{2Jy_2356_f}). There are no radio maps for PKS 1814$-$63 and PKS 1934$-$63, as these sources lack extended radio structures. FRI and FRII stand for Fanaroff-Riley class I and II respectively \citep{FR1974}; CSS and C/J stand for compact steep-spectrum and compact/jet, respectively; BL-LAC stands for BL Lacertae object. In terms of their nuclear (AGN) properties, LERG stands for low-excitation radio galaxy \citep[see e.g.][]{Laing1994}, NLRG and BLRG for narrow-line and broad-line radio galaxy, and Q for quasar. The references for the radio maps are: (1) \citet{Leahy1997}; (2) made directly from Karl G. Jansky Very Large Array (VLA) archive data; (3) \citet{Morganti1993}; (4) \citet{Morganti1999}; (5) \citet{Hardcastle2007}; (6) \citet{Gizani2003}; (7) \citet{Dennett-Thorpe2002}; (8) made directly from new Jansky Very Large Array (JVLA) data; (9) made directly from Australia Telescope Compact Array (ATCA) archive data.}}\label{2Jy_objects_table}
\centering
\setlength{\tabcolsep}{1.6pt}
\setlength{\extrarowheight}{2.6pt}
\begin{tabular}{cccccccccccc}\hline
PKS&3C&FR Class&AGN type&$z$&\textit{Chandra} obsid&Exp. time&Radio map freq.&Resolution&Peak flux&RMS&Ref.\\
&&&&&&(ks)&(GHz)&(arcsec)&mJy/beam&mJy/beam&\\\hline
0034$-$01&15&FRII&LERG&0.073&02176&28.18&8.4&$0.3\times0.3$&27.954&0.064&1\\
0038$+$09&18&FRII&BLRG&0.188&09293&8.05&4.9&$4.4\times3.4$&121.00&0.13&3\\
0043$-$42&&FRII&LERG&0.116&10319&18.62&8.6&$1.2\times0.88$&154.95&0.21&4\\
0213$-$13&62&FRII&NLRG&0.147&10320&20.15&4.9&$5.9\times3.4$&313.36&0.18&3\\
0349$-$27&&FRII&NLRG&0.066&11497&20.14&1.5&$11.0\times8.9$&876.96&0.18&2\\
0404$+$03&105&FRII&NLRG&0.089&09299&8.18&8.4&$2.2\times2.2$&384.90&0.17&1\\
0442$-$28&&FRII&NLRG&0.147&11498&20.04&4.9&$1.0\times0.6$&243.66&0.12&2\\
0521$-$36&&C/J&BL-LAC/BLRG&0.055&00846&9.87&4.7&$1.2\times0.7$&3022.6&1.8&3\\
0620$-$52&&FRI&LERG&0.051&11499&20.05&4.9&$2.6\times1.5$&237.711&0.046&3\\
0625$-$35&&FRI&LERG&0.055&11500&20.05&4.9&$4.7\times3.2$&690.63&0.14&2\\
0625$-$53&&FRI&LERG&0.054&04943&18.69&4.8&$2.0\times1.6$&23.61&0.94&4\\
0806$-$10&195&FRII&NLRG&0.110&11501&20.04&4.9&$2.4\times1.6$&119.689&0.075&2\\
0915$-$11&218&FRI&LERG&0.054&04969&98.2&1.4&$2.0\times1.5$&1257.28&0.64&2\\
&&&&&04970&100.13&&&&&\\
0945$+$07&227&FRII&BLRG&0.086&06842&30.17&1.5&$4.0\times4.0$&186.404&0.095&5\\
&&&&&07265&20.11&&&&&\\
1559+02&327&FRII&NLRG&0.104&06841&40.18&8.5&$2.2\times2.2$&23.871&0.021&5\\
1648$+$05&348&FRI&LERG&0.154&05796&48.17&1.5&$1.4\times1.4$&13.185&0.062&6\\
&&&&&06257&50.17&&&&&\\
1733$-$56&&FRII&BLRG&0.098&11502&20.12&4.7&$2.2\times1.9$&577.42&0.41&3\\
1814$-$63&&CSS&NLRG&0.063&11503&20.13&-&-&-&-&-\\
1839$-$48&&FRI&LERG&0.112&10321&20.04&4.7&$2.6\times1.7$&126.087&0.090&3\\
1934$-$63&&CSS&NLRG&0.183&11504&20.05&-&-&-&-&-\\
1949$+$02&403&FRII&NLRG&0.059&02968&50.13&1.5&$4.5\times4.1$&505.11&0.12&7\\
1954$-$55&&FRI&LERG&0.060&11505&20.92&4.8&$2.4\times1.3$&93.95&0.56&4\\
2135$-$14&&FRII&Q&0.200&01626&15.13&4.9&$5.5\times3.4$&211.02&0.15&3\\
2211$-$17&444&FRII&LERG&0.153&11506&20.04&1.5&$2.3\times1.5$&38.254&0.039&8\\
2221$-$02&445&FRII&BLRG&0.057&07869&46.20&8.2&$2.4\times2.4$&77.74&0.24&1\\
2356$-$61&&FRII&NLRG&0.096&11507&20.05&1.5&$7.2\times6.9$&1348.0&1.3&9\\\hline
\end{tabular}
\end{table*}

\subsection{The sample}

\edit{Table \ref{2Jy_objects_table} gives details of the 2Jy sample used in this paper. As in \cite{Mingo2014} and \cite{Hardcastle2006,Hardcastle2009}, we classify sources as LERGs based on their [OIII] equivalent widths, after the definition of \citet{Laing1994}, and on inspection of their optical spectra. This definition is consistent with the WLRG (weak line radio galaxy) classification, also often used in the literature to refer to these sources \citep[e.g.][]{Tadhunter1998,Buttiglione2009,Dicken2014}.}

\edit{In terms of their Fanaroff-Riley classification \citep{FR1974}, our 2Jy sample has 7 FRI, 16 FRII, and 3 compact sources. We have listed these classifications, as well as the AGN types, in Table \ref{2Jy_objects_table}.}

\edit{It is worth mentioning again that the 2Jy sample does not overlap with the 3CRR catalogue, due to the different location of the sources (the 3CRR catalogue covers sources in the Northern hemisphere, with $\delta>+10^{\circ}$). Some of the brightest 2Jy sources are included in the original 3C catalogue, as is the case for e.g.\ the quasar 3C\,273 (PKS 1226+02). Because the 2Jy selection was made at a higher frequency than the 3CRR sample, overall, more beamed sources are selected for the 2Jy sample than they are for the 3CRR, despite the steep-spectrum cut. Some of the implications of this fact are discussed in \citetalias{Mingo2014}.}

\subsection{X-ray analysis}

\edit{As mentioned in the previous Section, for the X-rays we analysed \textit{Chandra} observations for the low-$z$ sources in our sample, also listed in Table \ref{2Jy_objects_table}. Four low-$z$ sources (PKS 0404+03, 1814$-$63, 2135$-$14, 2221$-$02) have \textit{XMM} observations that we did not use, since the \textit{Chandra} images provided all the information needed for our analysis, and had a much better spatial resolution. Most of the observations were carried out at our request, using the ACIS-S CCD and no gratings; when using archival data we only considered ACIS-S and ACIS-I observations without gratings, and discarded calibration or very short observations that did not significantly contribute to the statistics. We reprocessed all the data presented by \citetalias{Mingo2014}, using {\sc ciao} 4.7 and the latest CALDB. We included the correction for VFAINT mode to minimise the issues with the background for all the sources with a count rate below 0.01 counts s$^{-1}$ and observed in VFAINT mode. While this correction is not essential to study the cores of the sources, it can improve the statistics for extended and faint emission, which we have analysed for this work.}

\subsection{Reduction and calibration of new radio data}\label{Reduction}

\edit{Most of our radio maps are taken from the work by \citet{Morganti1993,Morganti1999}. Table \ref{2Jy_objects_table} lists the radio map properties for each source, as well as the references for each dataset.}

\edit{A minority of sources were imaged afresh from VLA archive data. These were reduced in AIPS in the standard manner -- flux calibration used 3C\,48 or 3C\,286, a nearby point-source calibrator was used for phase calibration, and one or two iterations of phase followed by at most one iteration of amplitude self-calibration were carried out before final images were made at the full resolution of the data (using Briggs weighting with the robustness parameter set to 0). Where the structure of the source demanded it, data from different VLA configurations were combined and cross-calibrated before imaging. The one image made from archival ATCA data, that of PKS 2356$-$61, is composed of data from 3 different ATCA observations (in 3 different configurations) which were reduced in the standard manner in {\sc miriad} before being combined, self-calibrated and imaged in {\sc aips}.}

\edit{The data for PKS 2211$-$17 (3C\,444) are new broad-band (1-2 GHz) JVLA data obtained for a different purpose and will be discussed in more detail elsewhere (Mahatma et al.\ in prep.). For these data we used {\sc AOflagger} \citep{Offringa2012} on the raw data prior to data reduction to flag RFI. Data reduction was then performed on both A and B-configuration data sets individually, using {\sc casa} version 4.3.1, performed in the standard manner as described in the {\sc casa} tutorials \footnote{\url{https://casaguides.nrao.edu/index.php/Main_Page}}.}

\edit{For flux and bandpass calibration, 3C\,48 was observed in a single 3-minute scan. Phase and amplitude gain calibration was performed using the source J2246-1206. Bad baselines evident through the calibration process were flagged manually, as well as with the automated RFI flagging command `rflag'. The data were then averaged 16-fold so as to include 4 channels in each spectral window (16 spectral windows in total) with 512 MHz bandwidth per channel. Self-calibration was then performed in phase and amplitude on the individual A and B-configuration data sets before imaging both A and B-configuration data sets together, with a pixel size of $0.3\times0.3$ arcsec, and a clean noise threshold of 0.01 mJy.}

%

\section{The 2Jy sources}\label{Sources}

The following subsections briefly describe the images and spectra of the 2Jy sources imaged by \textit{Chandra}, with the exception of PKS 1226+02 (3C\,273), which was the first object to be identified as a quasar, and as such has been thoroughly studied in the past \citep[see e.g.][and references therein]{Soldi2008,Jester2005,Jester2006,Liu2011}.

All the X-ray images (Figs. \ref{2Jy_0034_f} to \ref{2Jy_2356_f}) correspond to ACIS-S observations, except for PKS 0625$-$53 and PKS 2135$-$14, which were taken with the ACIS-I. The images have been filtered to show just the 0.3--7 keV energy range, and are smoothed with a Gaussian profile with $\sigma=5$ pixels (1 pixel=0.492 arcsec), to better show the extended structures, except for PKS 0521$-$36 (Fig. \ref{2Jy_0521_f}), for which we used $\sigma=3$ pixels. 

\edit{Radio maps shown are listed in Table \ref{2Jy_objects_table}, where the peak flux and RMS for each map are also listed. No radio contours are shown in Figs. \ref{2Jy_1814_f} (PKS 1814$-$63) and \ref{2Jy_1934_f} (PKS 1934$-$63), since these are compact steep-spectrum sources (CSS) and have no extended radio structures. Although in this work we focus on X-ray emission from extended structures, we have also included images for these two compact sources, for completeness. For all the Figures, we have plotted radio contours uniformly covering the largest possible range of fluxes in each map, while also aiming to most clearly display the morphology of the sources, and avoid noise artifacts.}

\subsection{PKS 0034$-$01 (3C\,15)}\label{2Jy_0034-01}

\begin{figure}
\centering
\includegraphics[trim=7cm 5cm 7.5cm 4.5cm, clip=true, angle=0.0, width=0.47\textwidth]{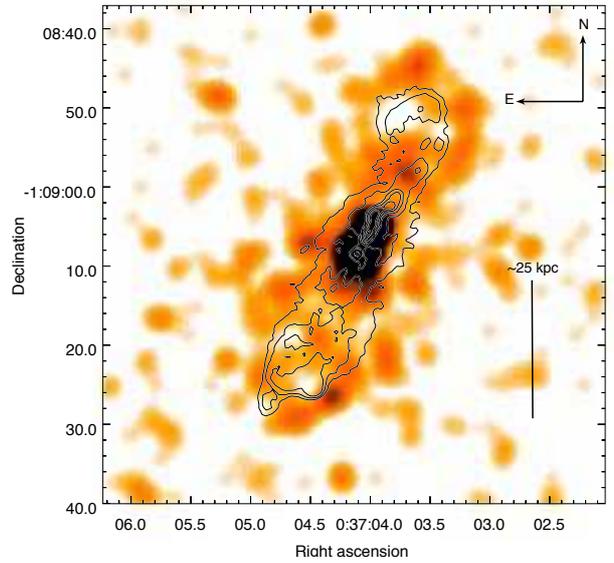}
\caption[PKS 0034-01]{PKS 0034$-$01 (3C\,15). The radio contours increase by factors of 2 between 0.0001 and 0.0128 Jy/beam, the beam major axis is 0.3 arcsec, and the minor axis is 0.3 arcsec.}\label{2Jy_0034_f}
\end{figure}

The radio morphology of PKS 0034$-$01 (Fig. \ref{2Jy_0034_f}) is intermediate between that of an FRI and an FRII, with a prominent jet in the N lobe but a weak hotspot in the S. The host galaxy sits in a relatively sparse environment, and it does not appear to be disturbed or interacting \citep[][]{RamosAlmeida2011}, showing no signs of recent star formation \citep{Dicken2012}, but it does have a dust lane \citep{Martel1999}. The \textit{Chandra} observation shows a 6 kpc ($\sim4$ arcsec) one-sided jet \citep[for a detailed study see][]{Dulwich2007}, which is also detected in radio \citep{Leahy1997,Morganti1999} and the Ks band \citep{Inskip2010}. There is also some X-ray emission coincident with the edges of the radio lobes \citep{Kataoka2003}, and its unusual X-ray nuclear emission has been discussed elsewhere (\citealt{VDWolk2010}; \citetalias{Mingo2014}). We have recently obtained new, deeper \textit{Chandra} data for 3C\,15, which will be presented in an upcoming paper.

\subsection{PKS 0038$+$09 (3C\,18)}\label{2Jy_0038+09}

\begin{figure}
\centering
\includegraphics[trim=4.5cm 2.0cm 5.0cm 2.5cm, clip=true, angle=0.0, width=0.47\textwidth]{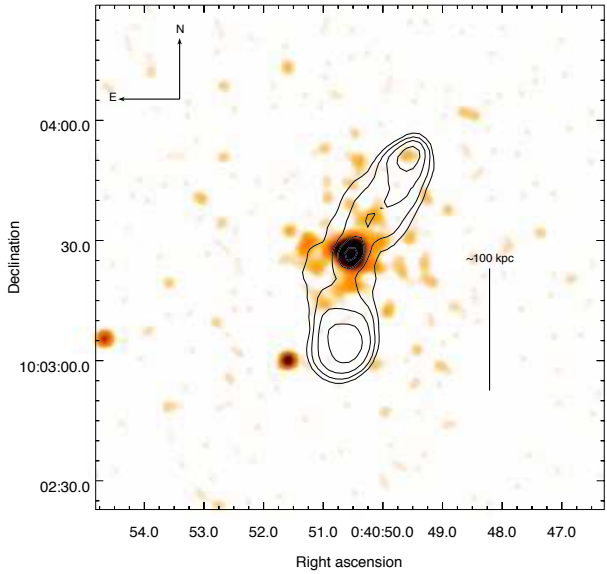}
\caption[PKS 0038+09]{PKS 0038$+$09 (3C\,18). The radio contours increase by factors of 2 between 0.007 and 0.112 Jy/beam, the beam major axis is 4.4 arcsec, and the minor axis is 3.4 arcsec.}\label{2Jy_0038_f}
\end{figure}

This BLRG seems to be in a dense environment, when observed in the optical \citep[][]{RamosAlmeida2013b}. We do not detect a luminous intracluster medium \citep[see also][]{Ineson2015}, but there seems to be some extended emission around the AGN in our images (Fig. \ref{2Jy_0038_f}). The X-ray image shows some enhanced emission coincident with the N hotspot, but the detection is not statistically significant ($1.5\sigma$), especially since there are similarly bright structures around it, so we have not included it in Table \ref{hotspots}.

\subsection{PKS 0043$-$42}\label{2Jy_0043-42}

\begin{figure}
\centering
\includegraphics[trim=7cm 5cm 7.5cm 4.5cm, clip=true, angle=0.0, width=0.47\textwidth]{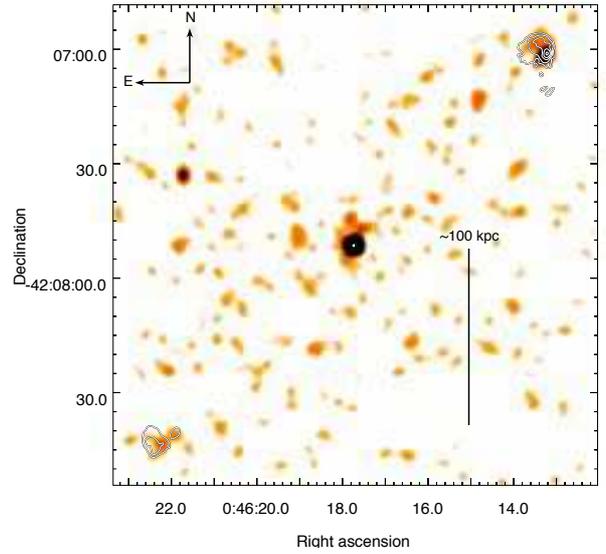}
\caption[PKS 0043-42]{PKS 0043$-$42. The radio contours increase by factors of 3 between 0.0015 and 0.1215 Jy/beam, the beam major axis is 1.3 arcsec, and the minor axis is 0.9 arcsec.}\label{2Jy_0043_f}
\end{figure}

Optical observations of PKS 0043$-$42 indicate that it inhabits a dense environment \citep{RamosAlmeida2013b}, from which we detect some faint extended emission in our \textit{Chandra} image (Fig. \ref{2Jy_0043_f}). \citet{Inskip2010} report a possible interaction with a nearby companion. \edit{Its radio morphology is very extended, and typical of a powerful FRII, with strong hotspots \citep{Morganti1999}}. We detect both hotspots in our X-ray image (see Table \ref{hotspots}), with a high significance in the case of the N hotspot ($5.3 \sigma$). It must be noted that, although this source is classified as a LERG, it shows signs of radiatively efficient accretion (see \citealt{RamosAlmeida2011b} and \citetalias{Mingo2014}).

\subsection{PKS 0213$-$13 (3C\,62)}\label{2Jy_0213-13}

\begin{figure}
\centering
\includegraphics[trim=7cm 5cm 7.5cm 4.5cm, clip=true, angle=0.0, width=0.47\textwidth]{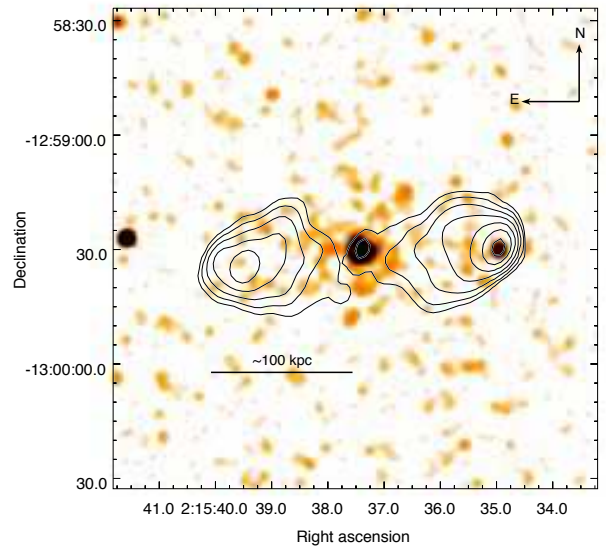}
\caption[PKS 0213-13]{PKS 0213$-$13 (3C\,62). The radio contours increase by factors of 2 between 0.003 and 0.192 Jy/beam, the beam major axis is 5.9 arcsec, and the minor axis is 3.4 arcsec.}\label{2Jy_0213_f}
\end{figure}

This NLRG has an optical shell and a narrow tidal tail \citep{RamosAlmeida2011}. The \textit{Chandra} image (Fig. \ref{2Jy_0213_f}) features a very bright hotspot W of the nucleus (Table \ref{hotspots}), in good agreement with the position of the radio emission. We do not detect the E hotspot in our X-ray image. We do detect an enhancement in emission inside the lobes, consistent with inverse-Compton scattering (see Section \ref{Lobes} and Table \ref{iCompton}).

\subsection{PKS 0349$-$27}\label{2Jy_0349-27}

\begin{figure}
\centering
\includegraphics[trim=6.5cm 4.5cm 7cm 4cm, clip=true, angle=0.0, width=0.47\textwidth]{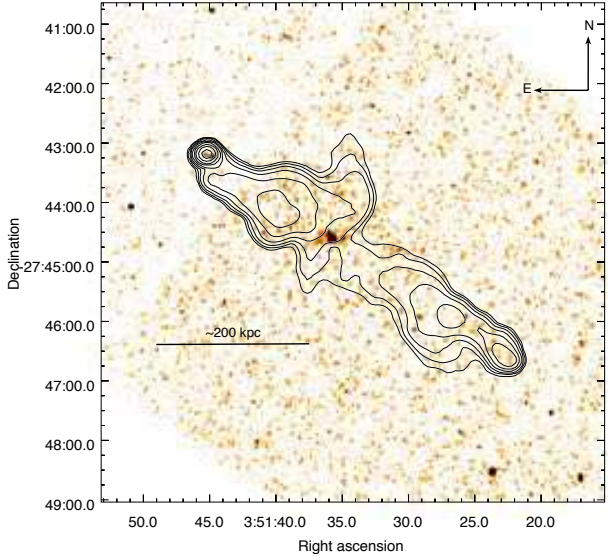}
\caption[PKS 0349-27]{PKS 0349$-$27. The radio contours increase by factors of 2 between 0.0016 and 0.4096 Jy/beam, the beam major axis is 11.0 arcsec, and the minor axis is 8.9 arcsec.}\label{2Jy_0349_f}
\end{figure}

This source is a well-known FRII galaxy, and it has some remarkable optical features, including an extended narrow line region and bridges connecting it to two neighbouring galaxies \citep{RamosAlmeida2013b,Inskip2010}, and an extended emission line nebulosity \citep{Danziger1984}. In our \textit{Chandra} image (Fig. \ref{2Jy_0349_f}) we detect some extended emission in the E-W direction, on scales of $\sim20$ kpc ($\sim16$ arcsec) around the nucleus, which could be associated with the optical bridges linking the host to the other galaxies or a hot medium. The emission towards the NE, in particular, along the expected direction of the jet, could correspond to the optical ionisation enhancement observed by \citet{Danziger1984}. We detect emission inside the lobes over the background level (see Section \ref{Lobes}), and observe an enhancement in emission with a slight offset ($\sim 6.4$ arcsec, equivalent to $\sim 8.3$ kpc) with the N radio hotspot (Table \ref{hotspots}, see also Fig. \ref{0349_hspotN}), although the offset may be partly caused by the fact that the X-ray emission falls very close to the edge of the CCD. We do not detect the S hotspot in X-rays.

\subsection{PKS 0404$+$03 (3C\,105)}\label{2Jy_0404+03}

\begin{figure}
\centering
\includegraphics[trim=6.5cm 4.5cm 7cm 4cm, clip=true, angle=0.0, width=0.47\textwidth]{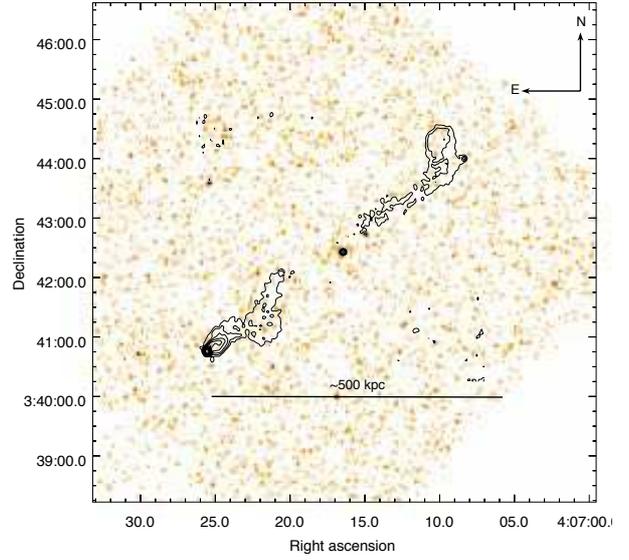}
\caption[PKS 0404+03]{PKS 0404$+$03 (3C\,105). The radio contours increase by factors of 2 between 0.0003 and 0.1536 Jy/beam, the beam major axis is 2.2 arcsec, and the minor axis is 2.2 arcsec.}\label{2Jy_0404_f}
\end{figure}

The host galaxy of PKS 0404$+$03 has been extensively studied in the IR and optical \citep[see][and references therein]{Inskip2010}, despite the high foreground $N_{H}$ column and the presence of a nearby star. The \textit{Chandra} image (Fig. \ref{2Jy_0404_f}) shows some emission coincident with the S radio hotspot (see Table \ref{hotspots}), which has also been studied in detail by \citet{Orienti2012}.

\subsection{PKS 0442$-$28}\label{2Jy_0442-28}

\begin{figure}
\centering
\includegraphics[trim=7.5cm 4.5cm 7.5cm 5cm, clip=true, angle=0.0, width=0.47\textwidth]{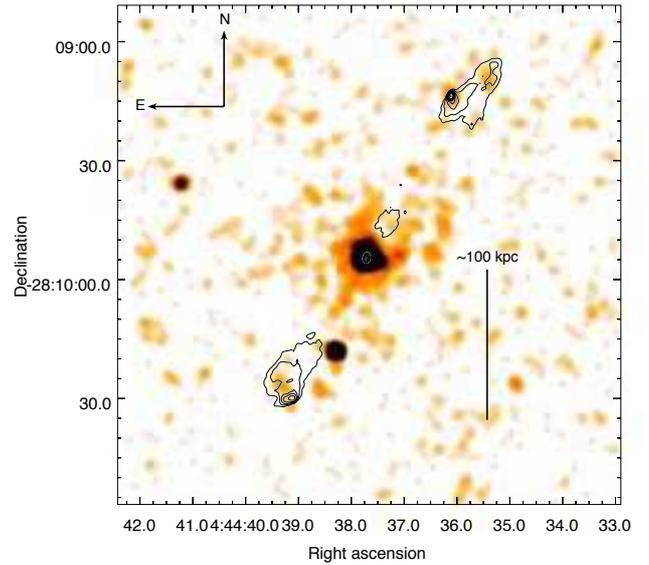}
\caption[PKS 0442-28]{PKS 0442$-$28. The radio contours increase by factors of 2 between 0.001 and 0.128 Jy/beam, the beam major axis is 1.0 arcsec, and the minor axis is 0.6 arcsec.}\label{2Jy_0442_f}
\end{figure}

The \textit{Chandra} image of this NLRG (Fig. \ref{2Jy_0442_f}) shows some extended emission, particularly surrounding the base of the N radio lobe. Although there is no ICM emission detected in the X-rays \citep{Ineson2015}, \citet{RamosAlmeida2013b} found several neighbouring galaxies. We also see a bright region coincident with the N hotspot, which we detect at a $3\sigma$ level (see Table \ref{hotspots}). We do not detect the S hotspot.

\subsection{PKS 0521$-$36}\label{2Jy_0521-36}

\begin{figure}
\centering
\includegraphics[trim=7cm 4.5cm 7.5cm 5cm, clip=true, angle=0.0, width=0.47\textwidth]{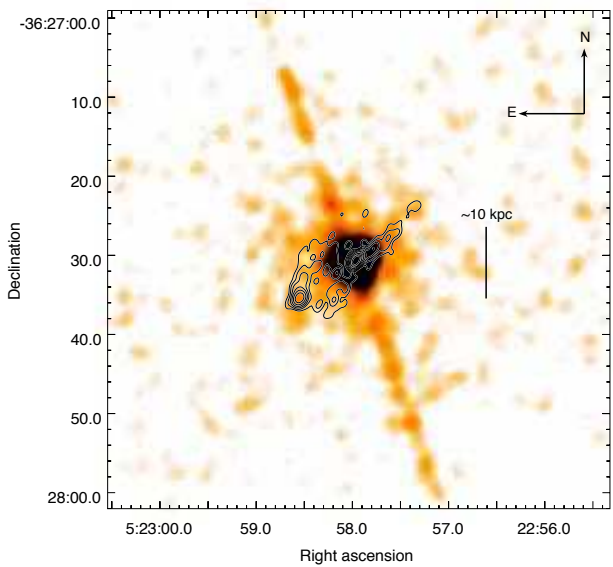}
\caption[PKS 0521-36]{PKS 0521$-$36. The radio contours increase by factors of 2 between 0.02 and 2.56 Jy/beam, the beam major axis is 1.3cm arcsec, and the minor axis is 0.7 arcsec. The instrumental streak is visible in the NE-SW direction.}\label{2Jy_0521_f}
\end{figure}

\edit{PKS 0521$-$36 is a very bright, misaligned BLRG with some peculiar spectral characteristics \citep[see][and references therein]{Inskip2010,DAmmando2015}, and an intermediate FRI/FRII structure.} The \textit{Chandra} image (Fig. \ref{2Jy_0521_f}) features a large streak, and is significantly piled up at the nucleus. \cite{Birkinshaw2002}, in their analysis of this dataset, report a detection of the core, jet, S hotspot and an extended, presumably thermal, halo.

\subsection{PKS 0620$-$52}\label{2Jy_0620-52}

\begin{figure}
\centering
\includegraphics[trim=7.5cm 4.5cm 7.5cm 5cm, clip=true, angle=0.0, width=0.47\textwidth]{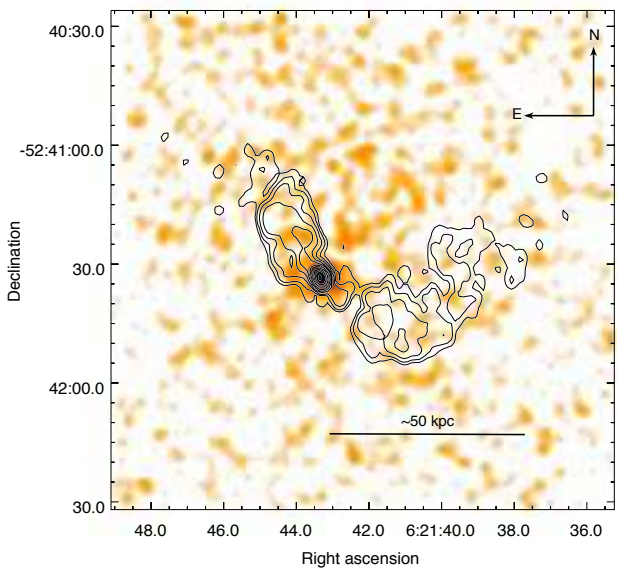}
\caption[PKS 0620-52]{PKS 0620$-$52. The radio contours increase by factors of 2 between 0.0005 and 0.1280 Jy/beam, the beam major axis is 2.6 arcsec, and the minor axis is 1.5 arcsec.}\label{2Jy_0620_f}
\end{figure}

This source has the lowest redshift in our sample, and it shows evidence for a young stellar population \citep{Dicken2012}. Although its optical morphology is not disturbed \citep{RamosAlmeida2011}, the presence of numerous nearby galaxies \citep{RamosAlmeida2013b}, and the fact that we detect extended emission in our \textit{Chandra} image \citep[Fig. \ref{2Jy_0620_f}, see also Fig. \ref{environments_f} for the larger-scale emission, and][]{Ineson2015}, make us agree with the hypothesis of \citet{Siebert1996}, \citet{Trussoni1999}, and \citet{Venturi2000} that this object sits in a rich cluster. The distorted shape of the radio lobes also indicates an interaction with the surrounding environment.

\subsection{PKS 0625$-$35}\label{2Jy_0625-35}

\begin{figure}
\centering
\includegraphics[trim=6cm 4cm 7cm 4.5cm, clip=true, angle=0.0, width=0.47\textwidth]{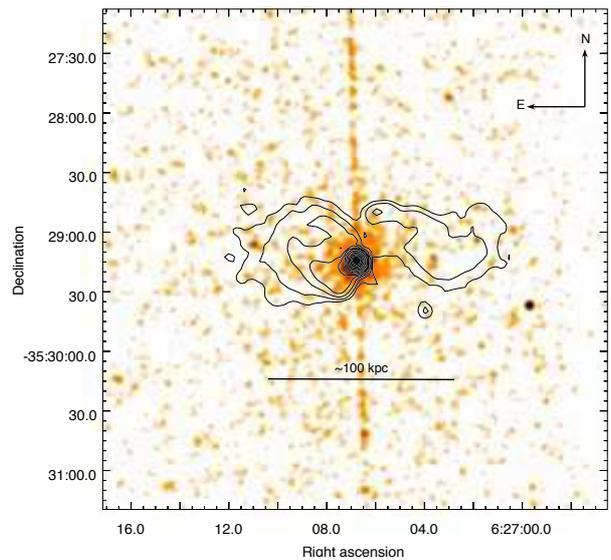}
\caption[PKS 0625-35]{PKS 0625$-$35. The radio contours increase by factors of 2 between 0.001 and 0.256 Jy/beam, the beam major axis is 4.7 arcsec, and the minor axis is 3.2 arcsec. The instrumental streak is visible in the N-S direction.}\label{2Jy_0625_35_f}
\end{figure}

PKS 0625$-$35 is suspected to be a BL Lac \citep{Wills2004}. It has a one-sided jet \citep{RamosAlmeida2011,Inskip2010}, which we do not resolve in the X-rays, and it does not seem to be interacting. The presence of a cluster environment was initially not clear \citep{Trussoni1999}, but it has recently been confirmed \citep{RamosAlmeida2013b,Ineson2015}. Although optically classified as a LERG, it is clear from our data that this is not a ``standard'' low-excitation object. The \textit{Chandra} image (Fig. \ref{2Jy_0625_35_f}) shows a large streak, and is piled up \citepalias[see also][]{Mingo2014}, but there are clear signs of a brightness gradient around the source (see Fig. \ref{environments_f}), indicating the possible presence of intra-cluster medium (ICM) emission from a dense environment.

\subsection{PKS 0625$-$53}\label{2Jy_0625-53}

\begin{figure}
\centering
\includegraphics[trim=7.0cm 4.5cm 7.5cm 5.0cm, clip=true, angle=0.0, width=0.47\textwidth]{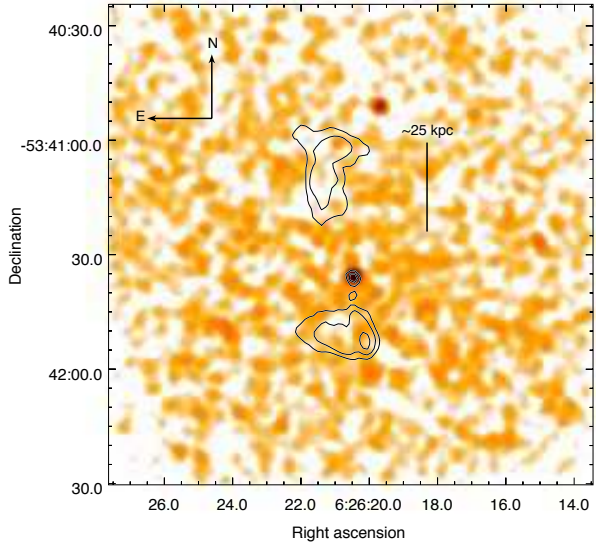}
\caption[PKS 0625-53]{PKS 0625$-$53. The radio contours increase by factors of 2 between 0.004 and 0.016 Jy/beam, the beam major axis is 2.0 arcsec, and the minor axis is 1.6 arcsec.}\label{2Jy_0625_53_f}
\end{figure}

PKS 0625$-$53 is a LERG hosted by a dumbbell galaxy, which is also the brightest member of the cluster Abell 3391 \citep{Frank2013,RamosAlmeida2013b,Ineson2015}. It has a `wide-angled tail' morphology \citep{Morganti1999} and a deflected jet. `Wide-angled tail' sources are traditionally classified as FRI, although they often show properties that are intermediate between both classes \citep[see e.g.][]{Hardcastle2004b,Jetha2006}. The optical images of PKS 0625$-$53 show a bridge of interaction with the W component of the dumbbell system \citep{RamosAlmeida2011}. The \textit{Chandra} image (Fig. \ref{2Jy_0625_53_f}) shows emission around the galaxy from the hot ICM, with a decrease in emission in the area overlapping with the N radio lobe, indicating a possible X-ray cavity.

\subsection{PKS 0806$-$10 (3C\,195)}\label{2Jy_0806-10}

\begin{figure}
\centering
\includegraphics[trim=7.0cm 4.5cm 7.5cm 5cm, clip=true, angle=0.0, width=0.47\textwidth]{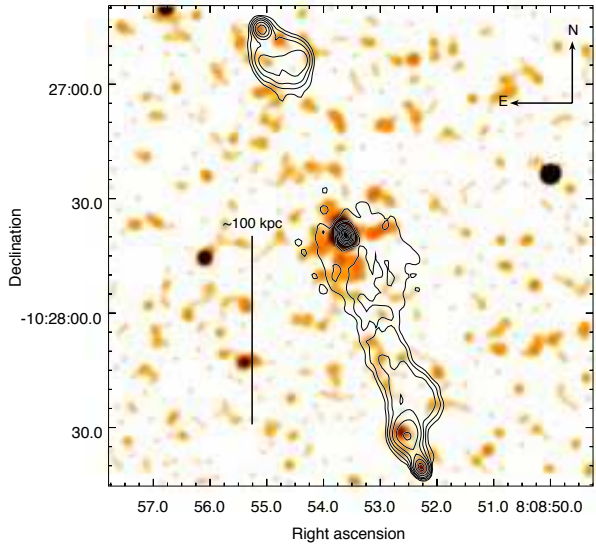}
\caption[PKS 0806-10]{PKS 0806$-$10 (3C\,195). The radio contours increase by factors of 2 between 0.001 and 0.064 Jy/beam, the beam major axis is 2.4 arcsec, and the minor axis is 1.6 arcsec.}\label{2Jy_0806_f}
\end{figure}

The optical and IR images of this galaxy show clear signs of disturbance \citep{Inskip2010,RamosAlmeida2011}. Our \textit{Chandra} image (Fig. \ref{2Jy_0806_f}) shows some enhancement in emission at the base of the radio lobes, near the nucleus, and enhancements in emission that are spatially coincident with the radio emission from the hotspots and S knot \citep{Morganti1993}. Around the N hotspot the X-ray emission is only enhanced at a $1.5\sigma$ level, with other structures of similar brightness around it, so we do not consider it a detection in Table \ref{hotspots}. We do detect the S hotspot and knot, however.

\subsection{PKS 0915$-$11 (3C\,218, Hydra A)}\label{2Jy_0915-11}

\begin{figure}
\centering
\includegraphics[trim=6cm 3.5cm 7cm 4.5cm, clip=true, angle=0.0, width=0.47\textwidth]{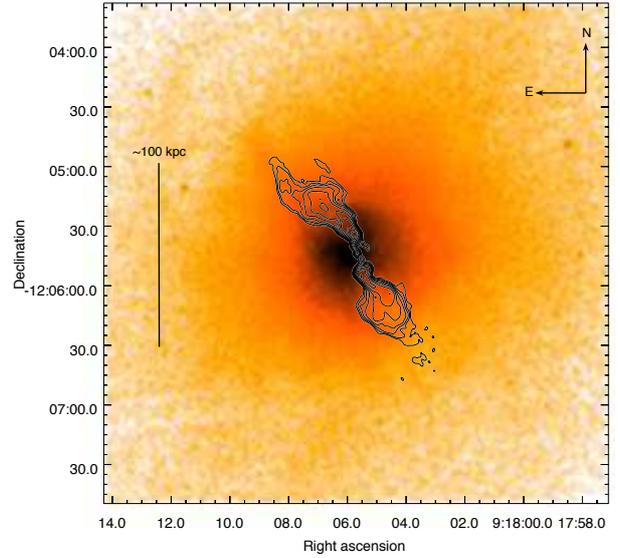}
\caption[PKS 0915-11]{PKS 0915$-$11 (3C\,218, Hydra A). The radio contours increase by factors of 2 between 0.004 and 1.024 Jy/beam, the beam major axis is 2.0 arcsec, and the minor axis is 1.5 arcsec.}\label{2Jy_0915_f}
\end{figure}

Hydra A is a very well-studied galaxy. It is one of the most powerful local radio sources, and it sits in the centre of a rich cluster \citep[see e.g.][and references therein]{Lane2004}. It shows evidence for recent star formation \citep{Dicken2012}, which is not common in cluster-centre galaxies, but can be attributed to a recent merger \citep[][report the presence of a dust lane]{RamosAlmeida2011}. The \textit{Chandra} images (Fig. \ref{2Jy_0915_f}) show the hot gas emission from the ICM, as well as emission associated with the lobes \citep[see e.g.][and references therein]{Kaastra2004,Wise2007,Hardcastle2010,Gitti2011}.

\subsection{PKS 0945$+$07 (3C\,227)}\label{2Jy_0945+07}

\begin{figure}
\centering
\includegraphics[trim=6.5cm 4cm 6cm 4.5cm, clip=true, angle=0.0, width=0.47\textwidth]{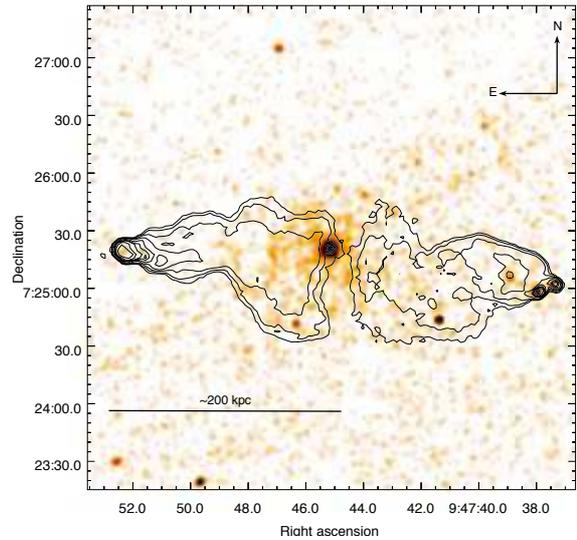}
\caption[PKS 0945+07]{PKS 0945$+$07 (3C\,227). The radio contours increase by factors of 2 between 0.0005 and 0.0128 Jy/beam, the beam major axis is 4.0 arcsec, and the minor axis is 4.0 arcsec. The instrumental streak is visible in the NW-SE direction.}\label{2Jy_0945_f}
\end{figure}

PKS 0945$+$07 is a well-known BLRG \citep{Morganti1993}, with a very extended optical emission line region \citep{Prieto1993}. The \textit{Chandra} image (Fig. \ref{2Jy_0945_f}) shows a faint readout streak. We detect some enhanced emission inside the radio lobes, whose spectrum is compatible with inverse-Compton scattering (see Section \ref{Lobes} and Table \ref{iCompton}), and bright X-ray emission coincident with the radio hotspots, particularly for the E structures \citep[see][for a detailed study of the hotspots, and also Table \ref{hotspots}]{Hardcastle2007}.

\subsection{PKS 1559$+$02 (3C\,327)}\label{2Jy_1559+02}

\begin{figure}
\centering
\includegraphics[trim=6.5cm 4cm 7cm 4.5cm, clip=true, angle=0.0, width=0.47\textwidth]{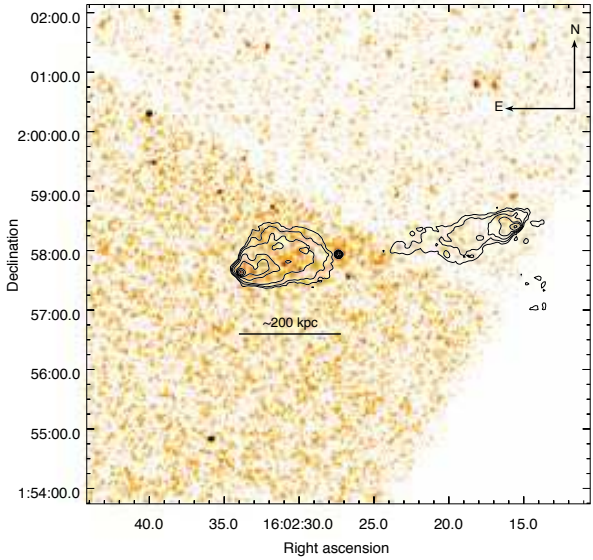}
\caption[PKS 1559+02]{PKS 1559$+$02 (3C\,327). The radio contours increase by factors of 2 between 0.0002 and 0.0256 Jy/beam, the beam major axis is 2.2 arcsec, and the minor axis is 2.2 arcsec.}\label{2Jy_1559_f}
\end{figure}

The host galaxy of this NLRG is very massive, and seems to have a bifurcated dust lane \citep{Inskip2010,RamosAlmeida2011}, which crosses the nucleus. Its radio morphology is extended and well known \citep{Morganti1993}, with the E lobe being much brighter than its W counterpart. \citet{VDWolk2010} report a large infrared excess that extends beyond what is expected for a torus. The \textit{Chandra} image (Fig. \ref{2Jy_1559_f}) shows a very bright nucleus, which is close to the edge of the S3 chip. As reported by \citet{Hardcastle2007}, there is enhanced emission within the E lobe (see Section \ref{Lobes} and Table \ref{iCompton}), with two bright spots coinciding with the E radio hotspot. \edit{It is worth mentioning that VLT observations show a foreground galaxy very close to the location of the E hotspot \citep{Mack2009}. }There seems to be some enhanced emission in the W lobe as well, but since it falls in one of the front-illuminated chips, and partly in the CCD gap, it is hard to quantify; we also do not detect a hotspot in the W lobe.

\subsection{PKS 1648$+$05 (3C\,348, Hercules A)}\label{2Jy_1648+05}

\begin{figure}
\centering
\includegraphics[trim=6cm 3.5cm 6.5cm 4.5cm, clip=true, angle=0.0, width=0.47\textwidth]{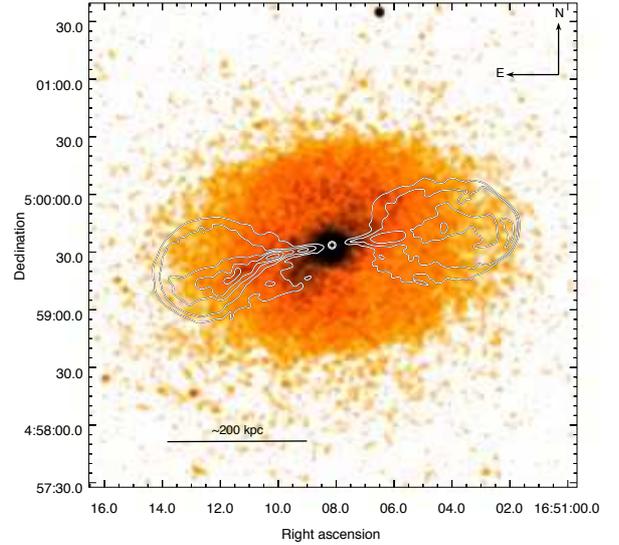}
\caption[PKS 1648+05]{PKS 1648$+$05 (3C\,348, Hercules A). The radio contours increase by factors of 3 between 0.002 and 0.162 Jy/beam, the beam major axis is 1.4 arcsec, and the minor axis is 1.4 arcsec.}\label{2Jy_1648_f}
\end{figure}

Hercules A is a cluster-embedded LERG with some unusual radio properties \citep{Morganti1993,Gizani2003}. Dust features are detected in the optical images \citep{RamosAlmeida2011}. The host galaxy is at the centre of a rich cluster \citep{RamosAlmeida2013b,Ineson2015}, and the lobes seem to be driving a shock into the ICM \citep{Nulsen2005,Nulsen2005b}, which is evident in the \textit{Chandra} image (Fig. \ref{2Jy_1648_f}), where there is clear emission from the hot ICM, with a lower density in the regions corresponding to the radio lobes. The nuclear X-ray spectrum is very faint \citepalias{Mingo2014}, with soft emission being the main contributor, as expected, and the X-ray images also show an enhancement in emission coincident with the radio jet, in the E direction. \cite{Hardcastle2010} have placed limits on the non-thermal emission associated with the lobes, but the extended emission is clearly dominated by thermal emission from the shocked ICM.

\subsection{PKS 1733$-$56}\label{2Jy_1733-56}

\begin{figure}
\centering
\includegraphics[trim=6cm 4cm 6.5cm 4.5cm, clip=true, angle=0.0, width=0.47\textwidth]{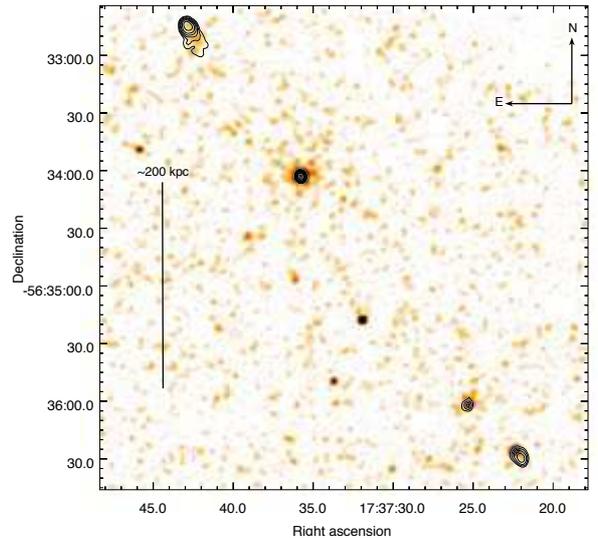}
\caption[PKS 1733-56]{PKS 1733$-$56. The radio contours increase by factors of 2 between 0.004 and 0.256 Jy/beam, the beam major axis is 2.2 arcsec, and the minor axis is 1.9 arcsec.}\label{2Jy_1733_f}
\end{figure}

The host galaxy of PKS 1733$-$56 shows evidence of recent star formation \citep{Dicken2012,Dicken2009}, and it has a disturbed optical morphology \citep{RamosAlmeida2011,Inskip2010}. Although there is a high foreground star density in the optical field, there are not many neighbouring galaxies near this source \citep{RamosAlmeida2013b}. The \textit{Chandra} image (Fig. \ref{2Jy_1733_f}) shows some diffuse emission, which could correspond to a hot ICM, and an enhancement in emission coincident with the radio hotspots. The N hotspot is the brighter in radio, but it is faint in X-rays, and there is extended emission around it, making its detection slightly unclear, (we have reported it on Table \ref{hotspots}, nonetheless, as statistically it is significant at a $3.2 \sigma$ level). We do detect, with high significance, the S hotspot and knot ($8.8\sigma$ and $6.9\sigma$, respectively), both of which are fainter in the radio. The knot is coincident with the radio emission, but the S hotspot seems slightly offset, by $\sim 5.5$ arcsec, corresponding to $\sim 10.2$ kpc (see also Fig. \ref{1733_hspotS}).

\subsection{PKS 1814$-$63}\label{2Jy_1814-63}

\begin{figure}
\centering
\includegraphics[trim=4cm 10cm 6.5cm 8.5cm, clip=true, width=0.47\textwidth]{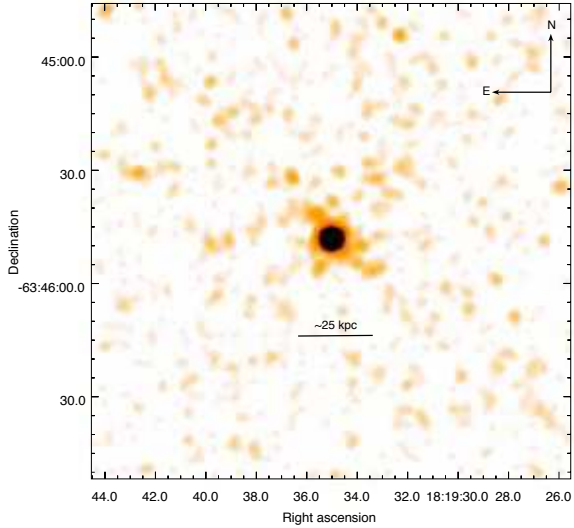}
\caption[PKS 1814-63]{PKS 1814$-$63. This source has no extended radio structures.}\label{2Jy_1814_f}
\end{figure}

PKS 1814$-$63 is a compact steep-spectrum radio source, and hence its core is not resolved by \textit{Chandra} (Fig. \ref{2Jy_1814_f}). The galaxy shows clear traces of an optical disk and a dust lane \citep{Inskip2010,RamosAlmeida2011}, which is atypical for a system with this radio luminosity \citep{Morganti2011}. It also shows evidence for starburst activity \citep{Dicken2012} and it has an extended emission line region \citep{Holt2008,Holt2009}. The \textit{Chandra} image shows no large-scale emission enhancement corresponding to a hot ICM, but there could be some extended emission near the AGN.

\subsection{PKS 1839$-$48}\label{2Jy_1839-48}

\begin{figure}
\centering
\includegraphics[trim=7cm 4.5cm 7.5cm 5cm, clip=true, angle=0.0, width=0.47\textwidth]{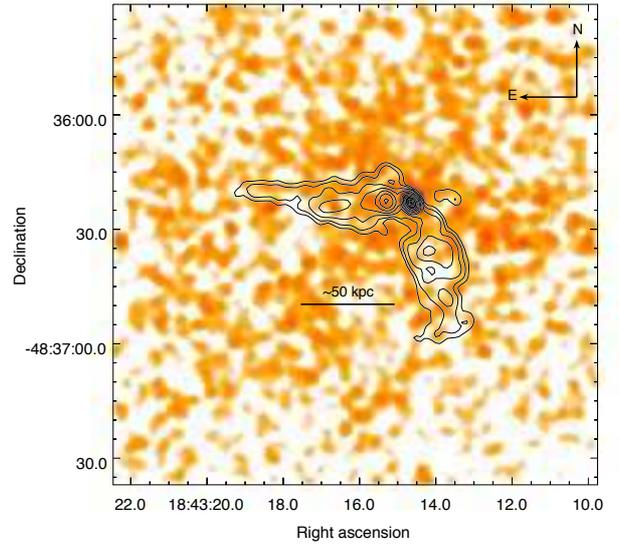}
\caption[PKS 1839-48]{PKS 1839$-$48. The radio contours increase by factors of 2 between 0.001 and 0.064 Jy/beam, the beam major axis is 2.6 arcsec, and the minor axis is 1.7 arcsec.}\label{2Jy_1839_f}
\end{figure}

This FRI is another example of a cluster-embedded LERG \citep{Tadhunter1993,RamosAlmeida2013b,Ineson2015}. Although not as dense as that of Hydra A or Hercules A, there is emission from the ICM in the \textit{Chandra} image (Fig. \ref{2Jy_1839_f}, see also Fig. \ref{environments_f}), and the radio lobes are clearly deflected by the interaction with the ICM, showing a `wide-angle tail' morphology.

\subsection{PKS 1934$-$63}\label{2Jy_1934-63}

\begin{figure}
\centering
\includegraphics[trim=4cm 10cm 6.5cm 8.5cm, clip=true, width=0.47\textwidth]{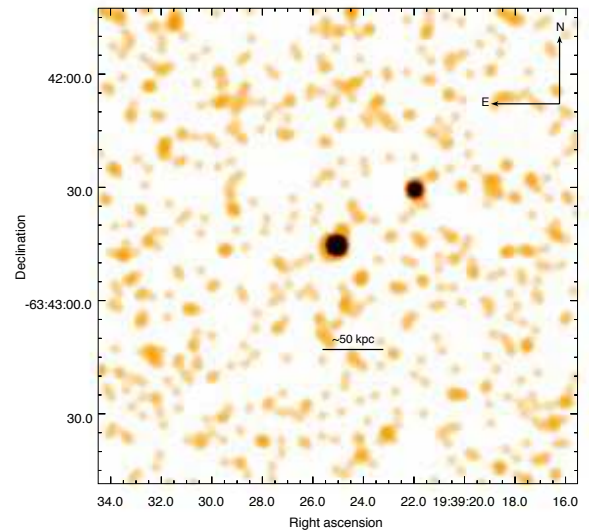}
\caption[PKS 1934-63]{PKS 1934$-$63. This source has no extended radio structures.}\label{2Jy_1934_f}
\end{figure}

This source has a compact radio structure \citep{Ojha2004b}, which is not resolved by \textit{Chandra} (Fig. \ref{2Jy_1934_f}). It is optically very blue \citep{RamosAlmeida2011}, as well as being part of an interacting galaxy pair \citep{Inskip2010}. It also shows evidence for infalling gas \citep{Holt2008,Holt2009}. The \textit{Chandra} image shows no signs of extended emission, only the compact source that coincides with the radio core.

\subsection{PKS 1949$+$02 (3C\,403)}\label{2Jy_1949+02}

\begin{figure}
\centering
\includegraphics[trim=6cm 4cm 7cm 4.5cm, clip=true, angle=0.0, width=0.47\textwidth]{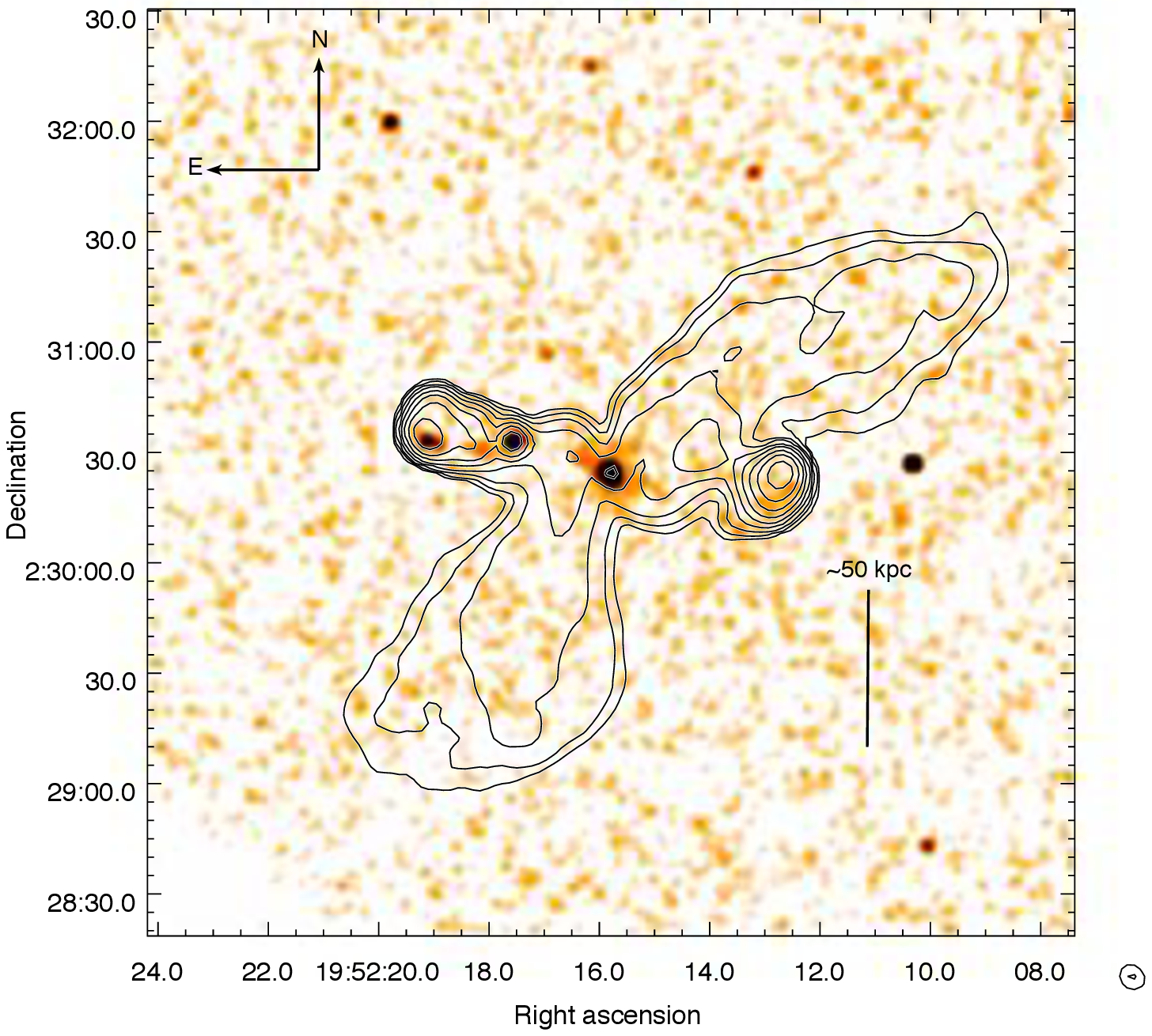}
\caption[PKS 1949+02]{PKS 1949$+$02 (3C\,403). The radio contours increase by factors of 2 between 0.001 and 0.256 Jy/beam, the beam major axis is 4.5 arcsec, and the minor axis is 4.1 arcsec.}\label{2Jy_1949_f}
\end{figure}

PKS 1949+02 is a NLRG with an X-shaped radio morphology, which has been studied in detail \citep[see][and references therein]{RamosAlmeida2011}. The \textit{Chandra} data have been studied in detail by \citet{Kraft2005}. They found the image (Fig. \ref{2Jy_1949_f}) to show some enhancement that could correspond to a dense medium, and two features to the E of the core (a hotspot and a knot) spatially coincident with the radio emission. There is also a bridge between both features, which might indicate emission from the jet, although it might also be hot gas. Some emission can also be observed close to the W radio hotspot, which is not detected in the X-rays.

\subsection{PKS 1954$-$55}\label{2Jy_1954-55}

\begin{figure}
\centering
\includegraphics[trim=6.5cm 4.5cm 7cm 5cm, clip=true, angle=0.0, width=0.47\textwidth]{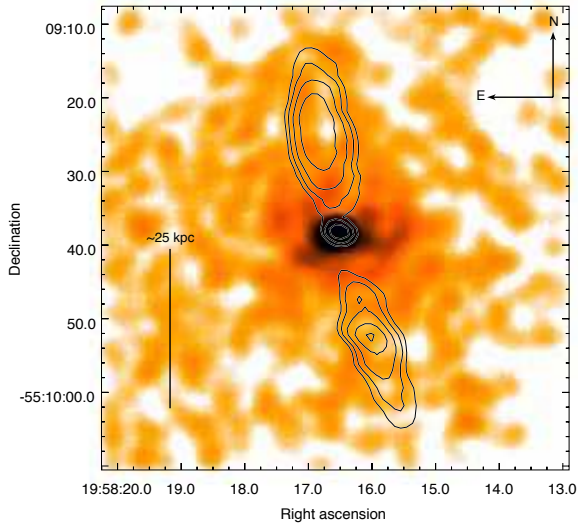}
\caption[PKS 1954-55]{PKS 1954$-$55. The radio contours increase by factors of 2 between 0.004 and 0.064 Jy/beam, the beam major axis is 2.4 arcsec, and the minor axis is 1.3 arcsec.}\label{2Jy_1954_f}
\end{figure}

PKS 1954$-$55 is another FRI LERG located the centre of a rich cluster \citep{RamosAlmeida2013b,Ineson2015}, whose hot gas emission is clearly visible in the X-rays (Fig. \ref{2Jy_1954_f}, see also Fig. \ref{environments_f}). The \textit{Chandra} image does not show clearly whether there are cavities associated with the lobes.

\subsection{PKS 2135$-$14}\label{2Jy2135-14}

\begin{figure}
\centering
\includegraphics[trim=6cm 4cm 7cm 4.5cm, clip=true, angle=0.0, width=0.47\textwidth]{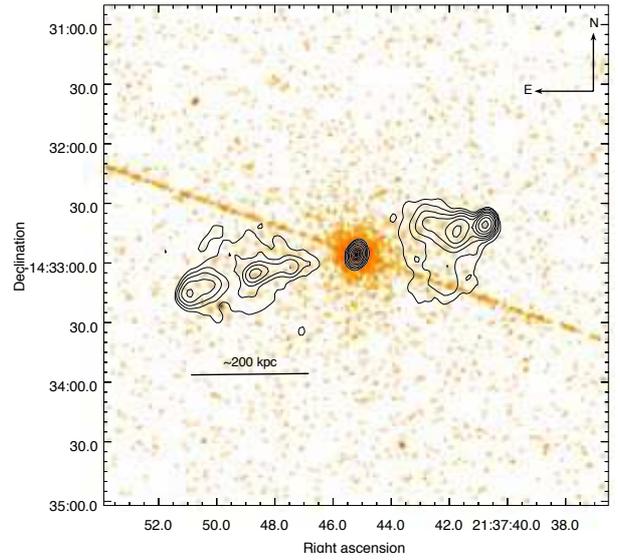}
\caption[PKS 2135-14]{PKS 2135$-$14. The radio contours increase by factors of 2 between 0.001 and 0.128 Jy/beam, the beam major axis is 5.5 arcsec, and the minor axis is 3.4 arcsec. The instrumental streak is visible in the NE-SW direction.}\label{2Jy_2135_f}
\end{figure}

The host of PKS 2135$-$14 has a close disk galaxy companion \citep{RamosAlmeida2011} and a disturbed morphology. The \textit{Chandra} image (Fig. \ref{2Jy_2135_f}) shows some extended emission around the nucleus, but given the brightness of this QSO (evidenced by the bright readout streak) it is difficult to tell whether that emission is from the PSF or a real ICM.

\subsection{PKS 2211$-$17 (3C\,444)}\label{2Jy_2211-17}

\begin{figure}
\centering
\includegraphics[trim=7cm 4cm 8cm 5cm, clip=true, angle=0.0, width=0.47\textwidth]{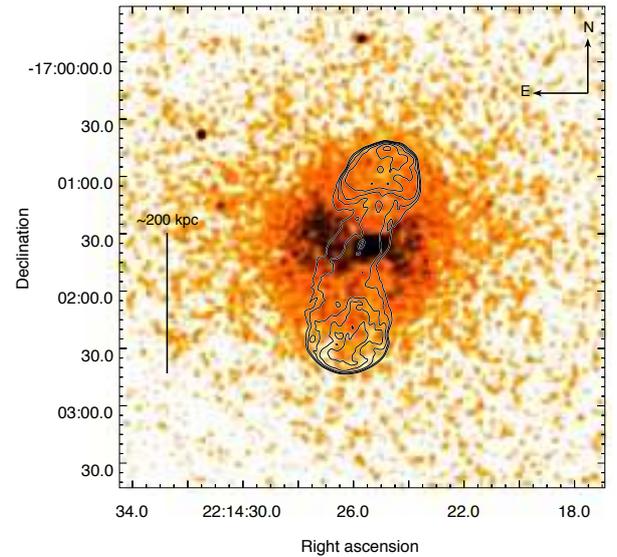}
\caption[PKS 2211-11]{PKS 2211$-$17 (3C\,444). The radio contours increase by factors of 2 between 0.001 and 0.032 Jy/beam, the beam major axis is 2.4 arcsec, and the minor axis is 1.3 arcsec.}\label{2Jy_2211_f}
\end{figure}

PKS 2211$-$17 (Fig. \ref{2Jy_2211_f}) is another cluster-embedded LERG \citep{Inskip2010,RamosAlmeida2013b,Ineson2015}. It is classified as an FRII, but its morphology is almost intermediate between the two FR classes. We detect a very dense ICM with clear cavities corresponding to the radio lobes, which are driving a shock \citep[][and in prep.]{Croston2011}. We used new 1.5 GHz JVLA radio data, processed by V. Mahatma as part of an on-going project, to generate the radio contours for Fig. \ref{2Jy_2211_f}.

\subsection{PKS 2221$-$02 (3C\,445)}\label{2Jy_2221-02}

\begin{figure}
\centering
\includegraphics[trim=6.5cm 4cm 7.5cm 4cm, clip=true, angle=0.0, width=0.47\textwidth]{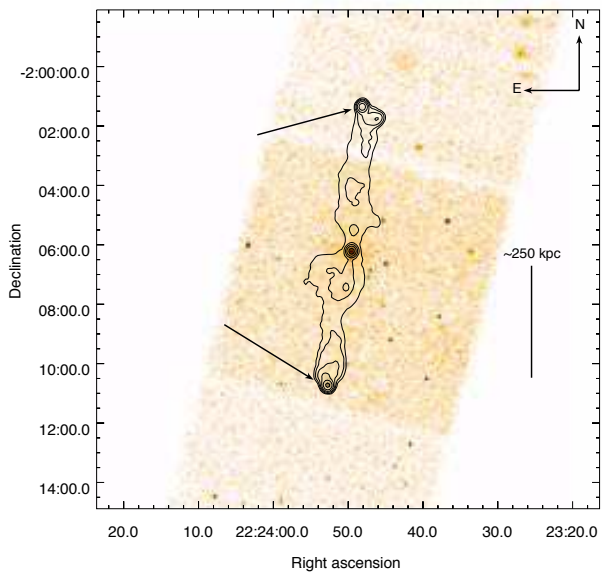}
\caption[PKS 2221-02]{PKS 2221$-$02 (3C\,445). The radio contours increase by factors of 2 between 0.001 and 0.032 Jy/beam, the beam major axis is 2.4 arcsec, and the minor axis is 2.4 arcsec. The arrows indicate the positions of the two hotspots.}\label{2Jy_2221_f}
\end{figure}

This object is a relatively well-known `double-double' BLRG \citep{Morganti1993,Leahy1997,Schoenmakers2000,Balmaverde2008,Inskip2010}. It seems to be interacting with a close companion \citep{RamosAlmeida2011}. The radio hotspots are detected by \textit{Chandra} (Fig. \ref{2Jy_2221_f}). The Northern one falls outside of the S3 chip, and it is not clearly detected, appearing at the $2.4 \sigma$ level \citep[see Table \ref{hotspots}, Fig. \ref{2221_hspotN}, and also][for a detailed analysis of the hotspots]{Orienti2012}, perhaps in part due to the slightly reduced sensitivity outside of the S3 chip. There seems to be some enhanced emission around the nucleus as well. Please note that, although we carried out all the analysis with the 8.2 GHz radio map of \citet{Leahy1997}, we used archival 4.9 GHz VLA radio data to generate the contours for Fig. \ref{2Jy_2221_f}, in order to show the large-scale radio lobes.

\subsection{PKS 2356$-$61}\label{2Jy_2356-61}

\begin{figure}
\centering
\includegraphics[trim=6cm 4cm 7cm 4cm, clip=true, angle=0.0, width=0.47\textwidth]{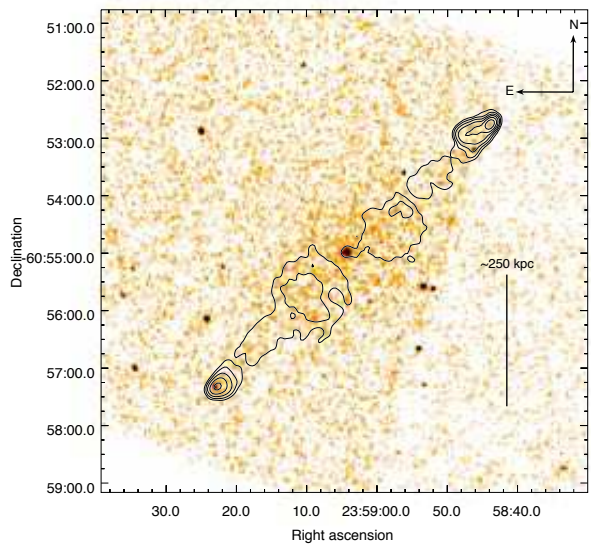}
\caption[PKS 2356-61]{PKS 2356$-$61. The radio contours increase by factors of 2 between 0.02 and 0.64 Jy/beam, the beam major axis is 7.2 arcsec, and the minor axis is 6.9 arcsec.}\label{2Jy_2356_f}
\end{figure}

The host of PKS 2356$-$61 shows signs of a past merger \citep{RamosAlmeida2011}. It is very radio powerful and has large hotspots and bright tails \citep{Subrahmanyan1996}, with the S hotspot being detected at a $6\sigma$ level in our \textit{Chandra} image (Fig. \ref{2Jy_2356_f}, see also Table \ref{hotspots}). Although there is some emission in the area around the N hotspot, we do not detect it. There is also X-ray inverse-Compton emission inside the lobes (Table \ref{iCompton}), and emission around the source and at the base of the lobes which could be related to a hot ICM.

%

\section{Hotspots}\label{Hspots}

\begin{table*}\small
\caption{Hotspots and knots. The photon index of the X-ray powerlaw ($\Gamma$) was fixed to 1.9 when the statistics did not allow a model fit; these fixed values are indicated with an asterisk. When no $\sim8.4$ GHz data were available, the radio fluxes for the hotspots were extrapolated from the frequency listed on Table \ref{2Jy_objects_table}, using a spectral index $\alpha=0.6$. See also Fig. \ref{hotspot_fig}, for details on the structures listed on this table. The results for PKS 0945$+$07 (3C\,227) and PKS 1559$+$02 (3C\,327) were extracted directly from \citet{Hardcastle2007}; the results for PKS 1949$+$02 (3C\,403) were extracted directly from \citet{Kraft2005}. The X-ray results for PKS 0521$-$36 were extracted from \citet{Birkinshaw2002}. Although the detections of the N lobes of PKS 1733$-$56 and PKS 2221$-02$ are slightly dubious (see sections \ref{2Jy_1733-56} and \ref{2Jy_2221-02}, and Fig. \ref{hotspot_fig}), we report them here for completeness. The errors quoted only take into account the flux measurement uncertainty, please see Table \ref{2Jy_objects_table} and the references listed therein for the RMS values and details on the individual calibration uncertainties.}\label{hotspots}
\centering
\setlength{\extrarowheight}{1pt}
\begin{tabular}{ccccccccc}\hline
PKS&3C&Structure&0.3--7 keV net counts&$\Gamma$&1 keV flux dens.&Radio freq.&Radio flux dens.&X-ray/Radio\\
&&&&&(nJy)&(GHz)&(mJy)&$\times10^{-9}$\\\hline
0043$-$42&&N hotspot&$32\pm6$&1.9*&$1.5\pm0.3$&8.5&$582\pm1$&2.6\\
&&S hotspot&$6\pm3$&1.9*&$0.3\pm0.2$&8.5&$119\pm1$&2.5\\
0213$-$13&62&W hotspot&$23\pm5$&1.9*&$1.2\pm0.3$&8.5&$226.5\pm0.4$&5.3\\
0349$-$27&&N hotspot&$31\pm8$&1.9*&$2.2\pm0.8$&4.8&$988.7\pm0.8$&2.2\\
0404$+$03&105&S hotspot&$22\pm5$&1.9*&$5.5\pm1.3$&8.4&$1031\pm1$&5.3\\
0442$-$28&&N hotspot&$9\pm3$&1.9*&$0.5\pm0.2$&4.9&$406\pm1$&1.2\\
0521$-$36&&E hotspot&$5\pm1$&1.9*&$0.4\pm0.1$&4.9&$1220\pm1$&0.3\\
0806$-$10&195&S knot&$7\pm3$&1.9*&$0.4\pm0.2$&4.9&$130.5\pm0.2$&3.1\\
&&S hotspot&$8\pm3$&1.9*&$0.5\pm0.2$&4.9&$151.0\pm0.2$&3.3\\
0945$+$07&227&E hotspot&$10\pm3$&1.6*&$0.3\pm0.1$&8.4&5.3&56.6\\
&&W primary hotspot&$84\pm9$&$1.6\pm0.2$&$1.5\pm0.2$&8.4&41&36.6\\
&&W secondary hotspot&$19\pm5$&1.6*&$0.3\pm0.1$&8.4&13&23.1\\
&&W knot&$22\pm5$&1.6*&$0.4\pm0.1$&8.4&1.2&333.3\\
1559$+$02&327&E primary hotspot&$12\pm3$&1.7*&$0.3\pm0.1$&8.4&15&18.0\\
&&E secondary hotspot&$12\pm3$&1.7*&$0.3\pm0.1$&8.4&3.2&84.4\\
1733$-$56&&N hotspot&$13\pm4$&1.9*&$0.8\pm0.2$&4.7&$1252\pm2$&0.6\\
&&S hotspot&$88\pm10$&$1.9\pm0.4$&$5.1\pm1.1$&4.7&$315\pm2$&16.2\\
&&S knot&$55\pm8$&$1.9\pm0.4$&$3.0\pm0.8$&4.7&$105\pm2$&28.6\\
1949$-$02&403&E hotspot&$44\pm6$&$1.8\pm0.4$&$0.9\pm0.2$&8.4&25&40.00\\
&&E knot&$83\pm9$&$1.7\pm0.3$&$2.3\pm0.2$&8.4&41&60.0\\
2221$-$02&445&N hotspot&$12\pm5$&1.9*&$0.3\pm0.1$&8.2&$87.8\pm0.9$&3.4\\
&&S hotspot&$174\pm14$&$2.0\pm0.2$&$3.9\pm0.6$&8.2&$126.3\pm0.9$&30.9\\
2356$-$61&&S hotspot&$61\pm10$&$1.9\pm0.5$&$3.0\pm0.5$&1.4&$1875\pm4$&1.6\\\hline
\end{tabular}
\end{table*}

\begin{figure*}
\centering
\begin{subfigure}[b]{0.24\linewidth}
\centering
\frame{\includegraphics[trim=7.5cm 4.5cm 7cm 4cm, clip=true, angle=0.0, width=.9\linewidth]{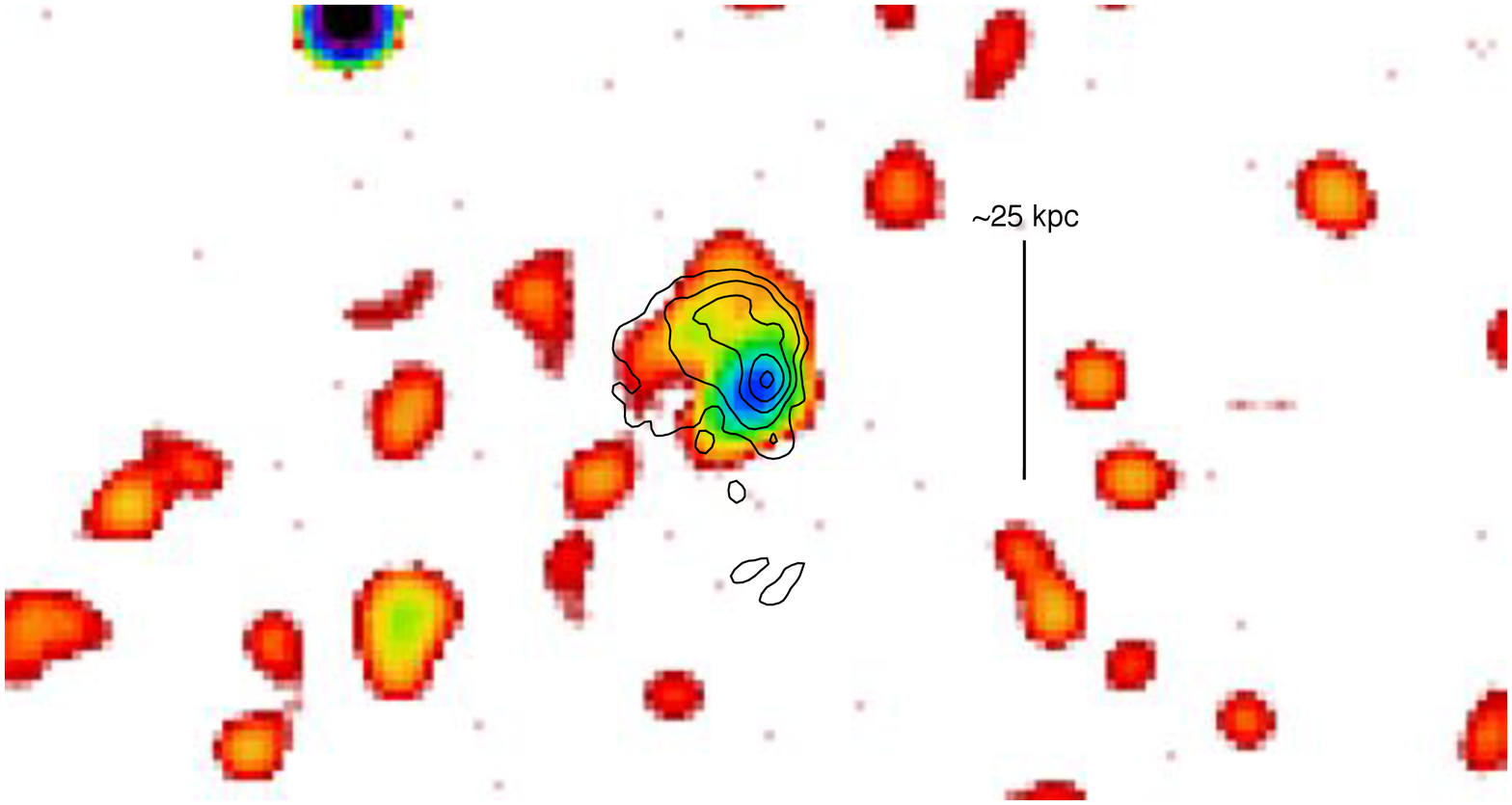}}
\caption{\small{PKS 0043$-$42, N hotspot}}\label{0043_hspotN}
\vspace{1ex}
\end{subfigure}
\begin{subfigure}[b]{0.24\linewidth}
\centering
\frame{\includegraphics[trim=7.5cm 4.5cm 7cm 4cm, clip=true, angle=0.0, width=.9\linewidth]{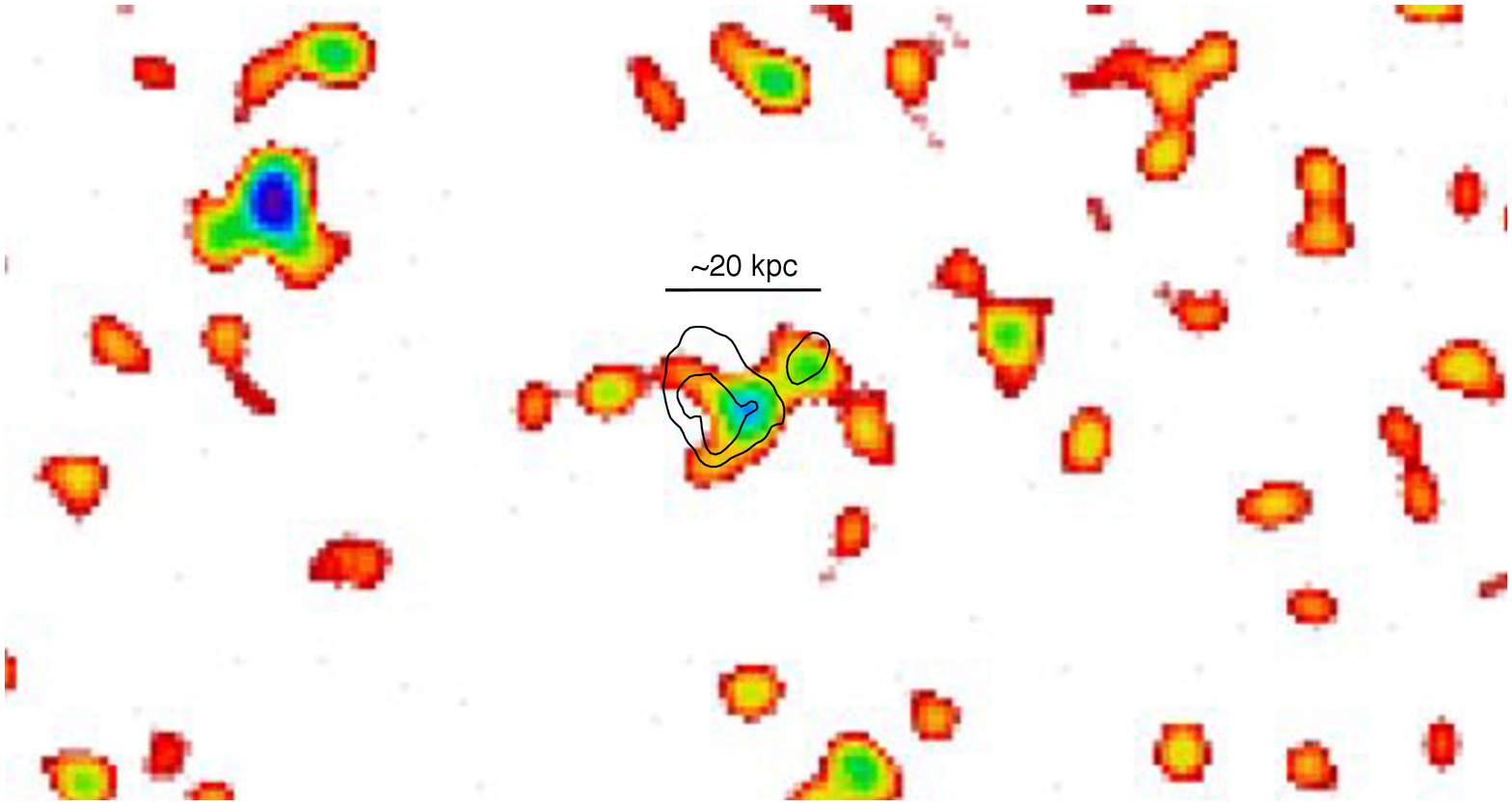}}
\caption{\small{PKS 0043$-$42, S hotspot}}\label{0043_hspotS}
\vspace{1ex}
\end{subfigure}
\begin{subfigure}[b]{0.24\linewidth}
\centering
\frame{\includegraphics[trim=7.5cm 4.5cm 7cm 4cm, clip=true, angle=0.0, width=.9\linewidth]{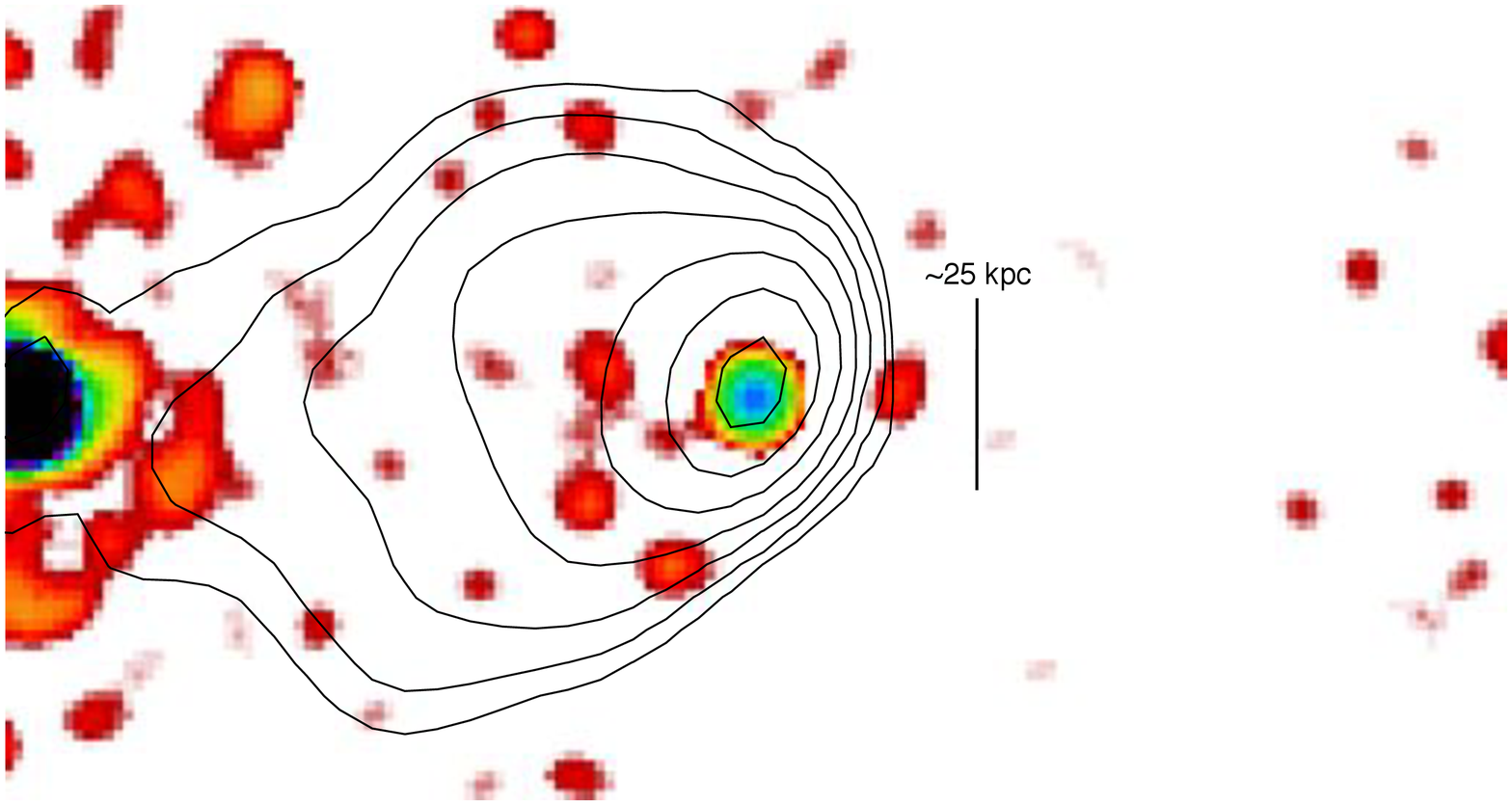}}
\caption{\small{PKS 0213$-$13, W hotspot}}\label{0213_hspotW}
\vspace{1ex}
\end{subfigure}
\begin{subfigure}[b]{0.24\linewidth}
\centering
\frame{\includegraphics[trim=7.5cm 4.5cm 7cm 4cm, clip=true, angle=0.0, width=.9\linewidth]{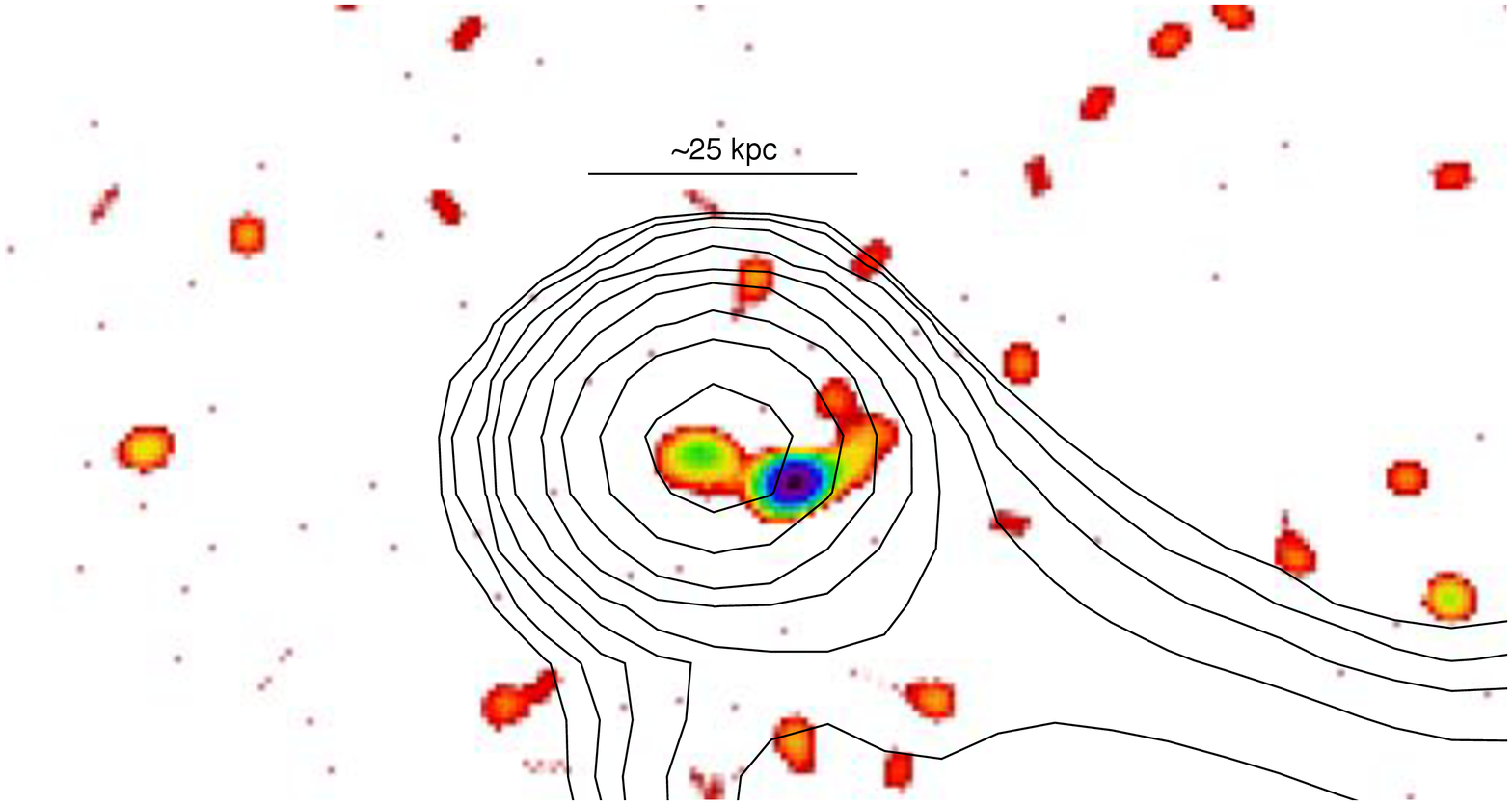}}
\caption{\small{PKS 0349$-$27, N hotspot}}\label{0349_hspotN}
\vspace{1ex}
\end{subfigure}
\begin{subfigure}[b]{0.24\linewidth}
\centering
\frame{\includegraphics[trim=7.5cm 4.5cm 7cm 4cm, clip=true, angle=0.0, width=.9\linewidth]{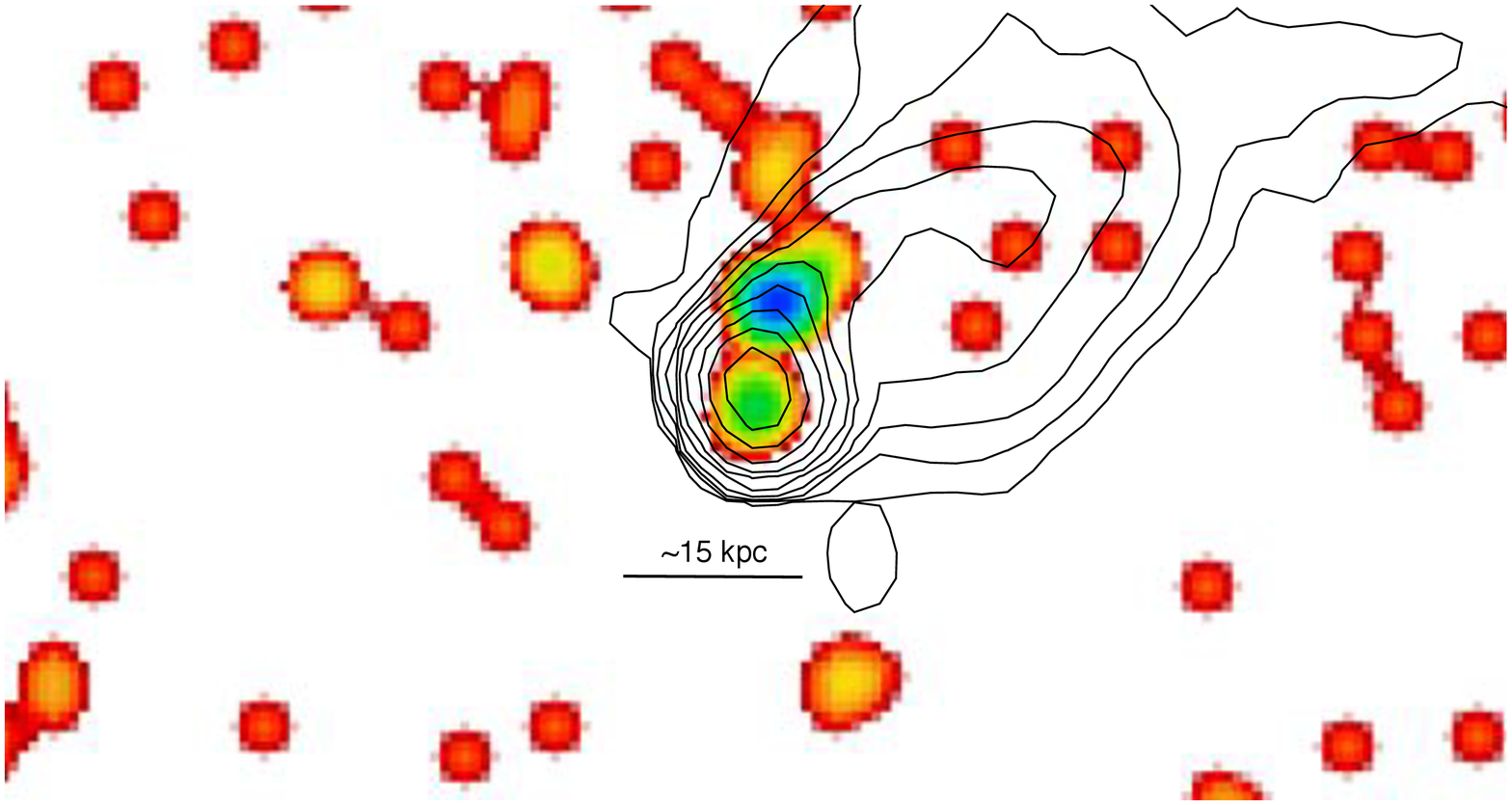}}
\caption{\small{PKS 0404$+$03, S hotspot}}\label{0404_hspotS}
\vspace{1ex}
\end{subfigure}
\begin{subfigure}[b]{0.24\linewidth}
\centering
\frame{\includegraphics[trim=7.5cm 4.5cm 7cm 4cm, clip=true, angle=0.0, width=.9\linewidth]{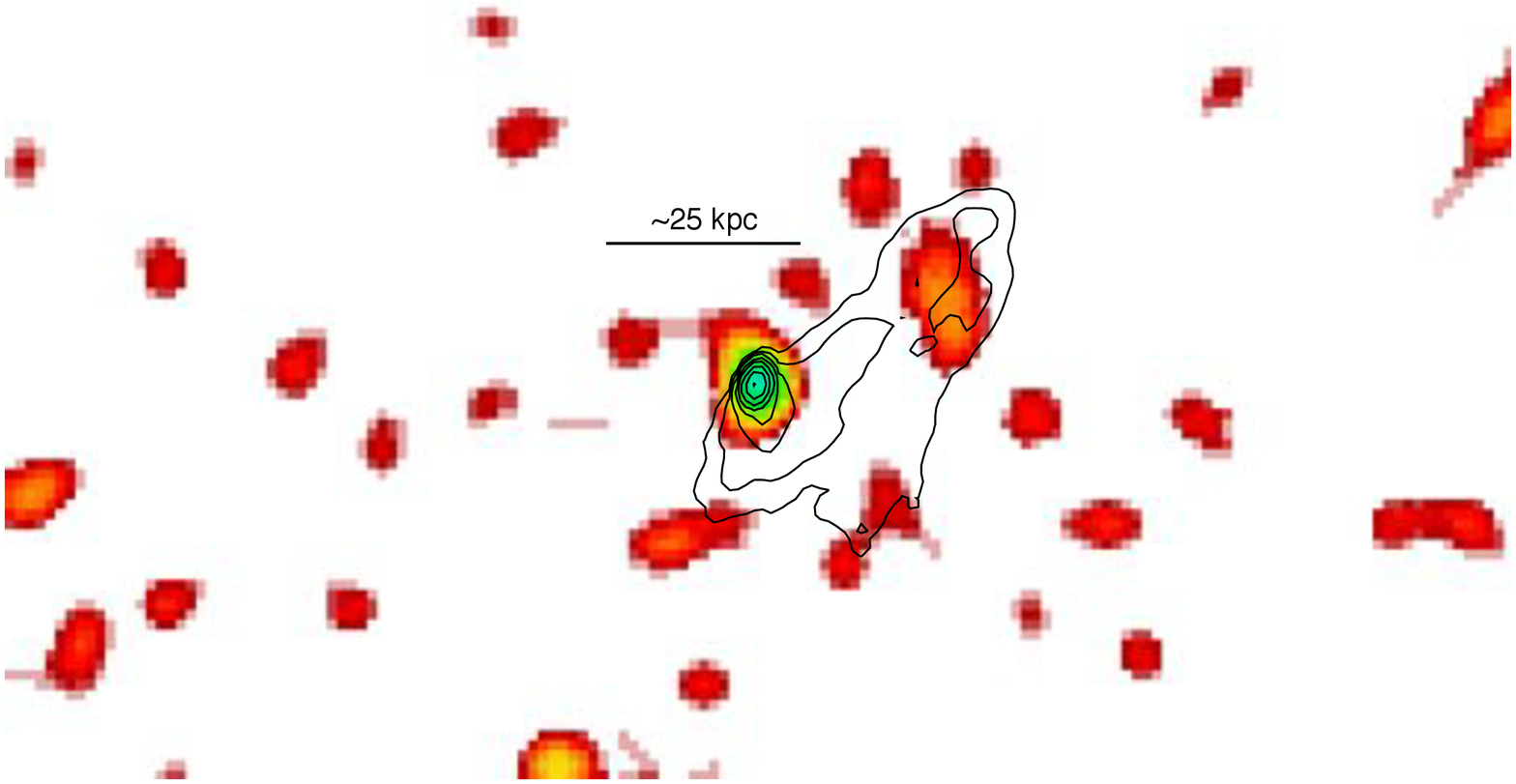}}
\caption{\small{PKS 0442$-$28, N hotspot}}\label{0442_hspotN}
\vspace{1ex}
\end{subfigure}
\begin{subfigure}[b]{0.24\linewidth}
\centering
\frame{\includegraphics[trim=7.5cm 4.5cm 7cm 4cm, clip=true, angle=0.0, width=.9\linewidth]{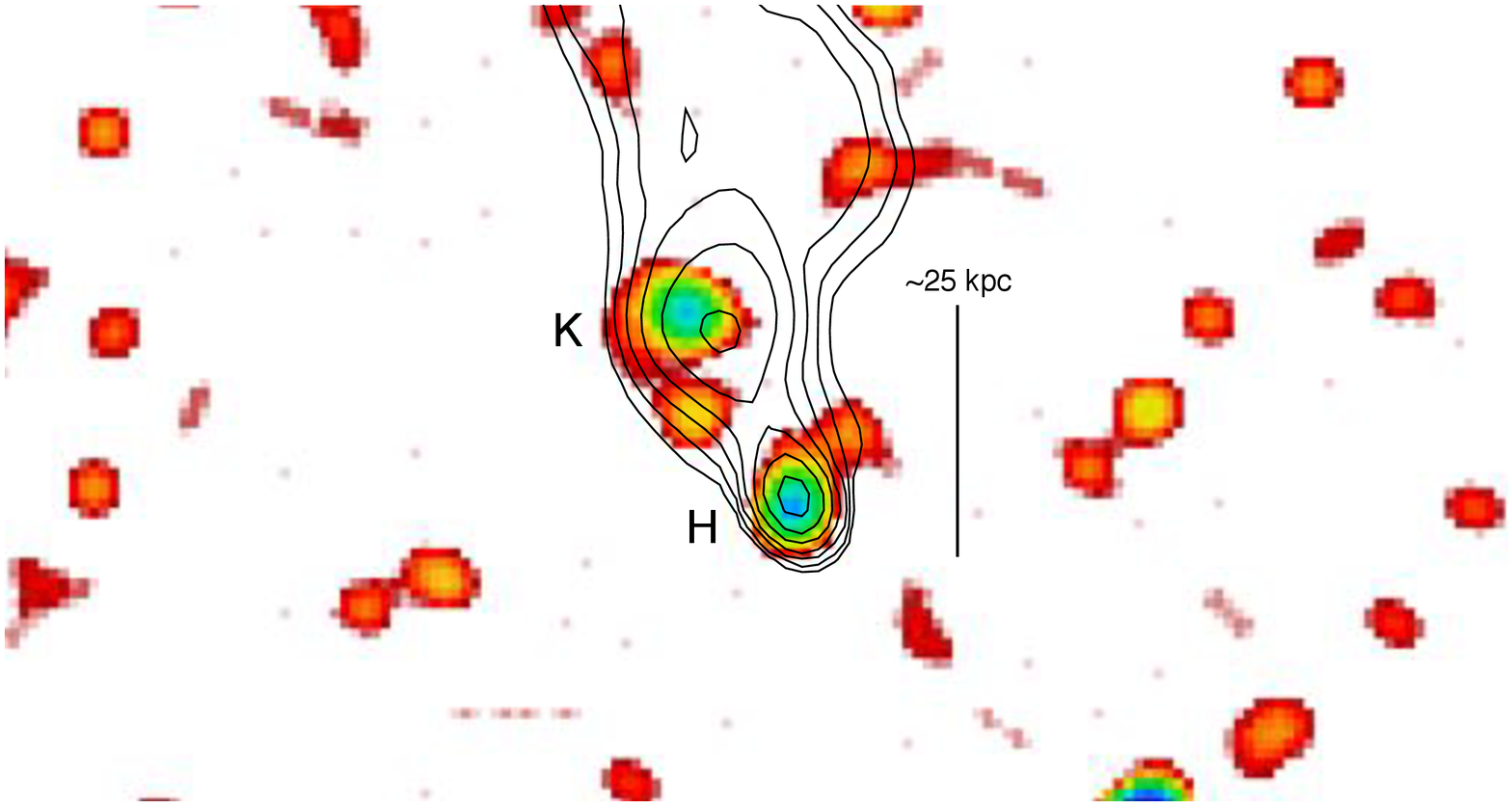}}
\caption{\small{PKS 0806$-$10, S knot and hotspot}}\label{0806_hspotS}
\vspace{1ex}
\end{subfigure}
\begin{subfigure}[b]{0.24\linewidth}
\centering
\frame{\includegraphics[trim=7.5cm 4.5cm 7cm 4cm, clip=true, angle=0.0, width=.9\linewidth]{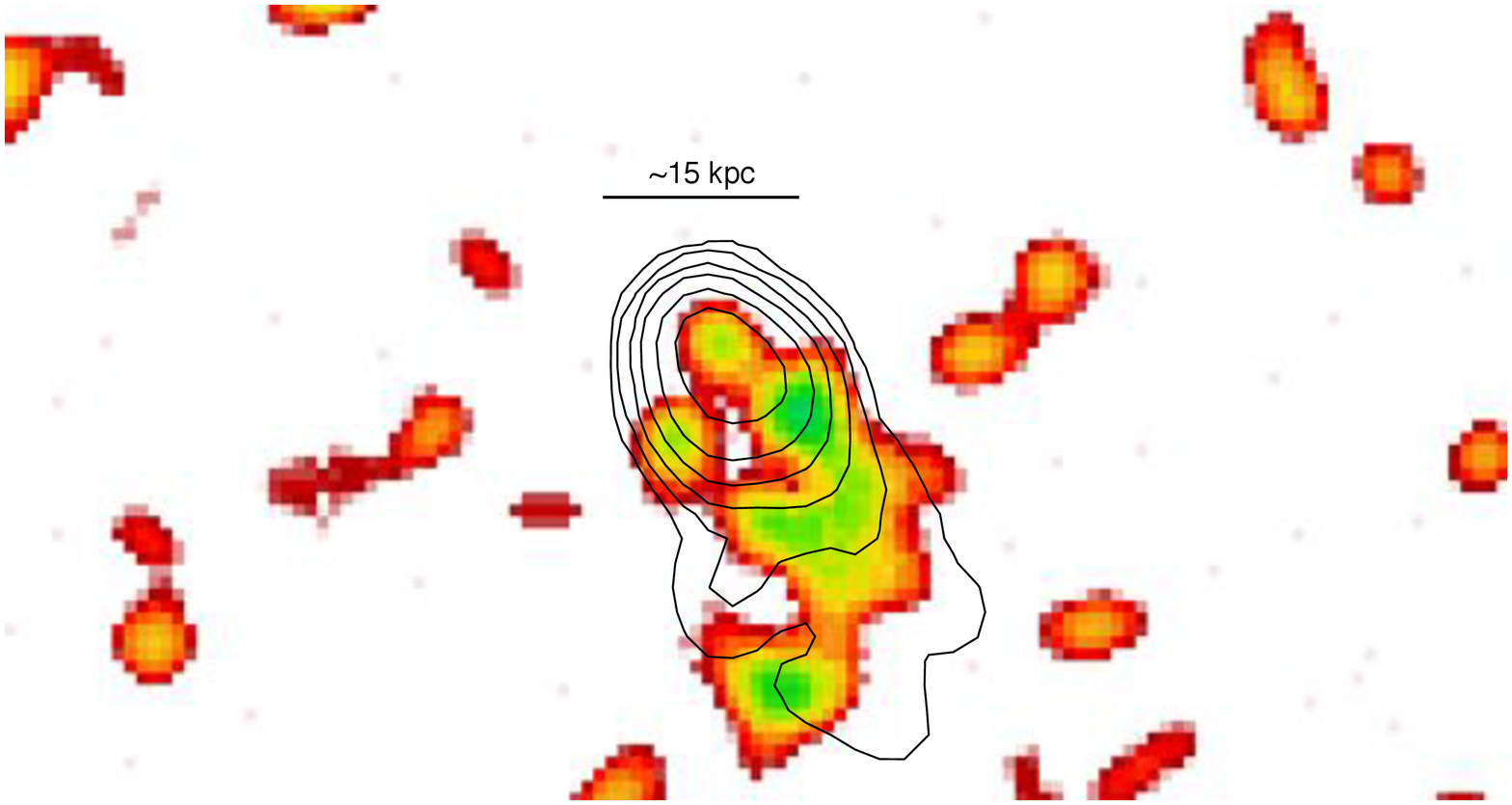}}
\caption{\small{PKS 1733$-$56, N hotspot}}\label{1733_hspotN}
\vspace{1ex}
\end{subfigure}
\begin{subfigure}[b]{0.24\linewidth}
\centering
\frame{\includegraphics[trim=7.5cm 4.5cm 7cm 4cm, clip=true, angle=0.0, width=.9\linewidth]{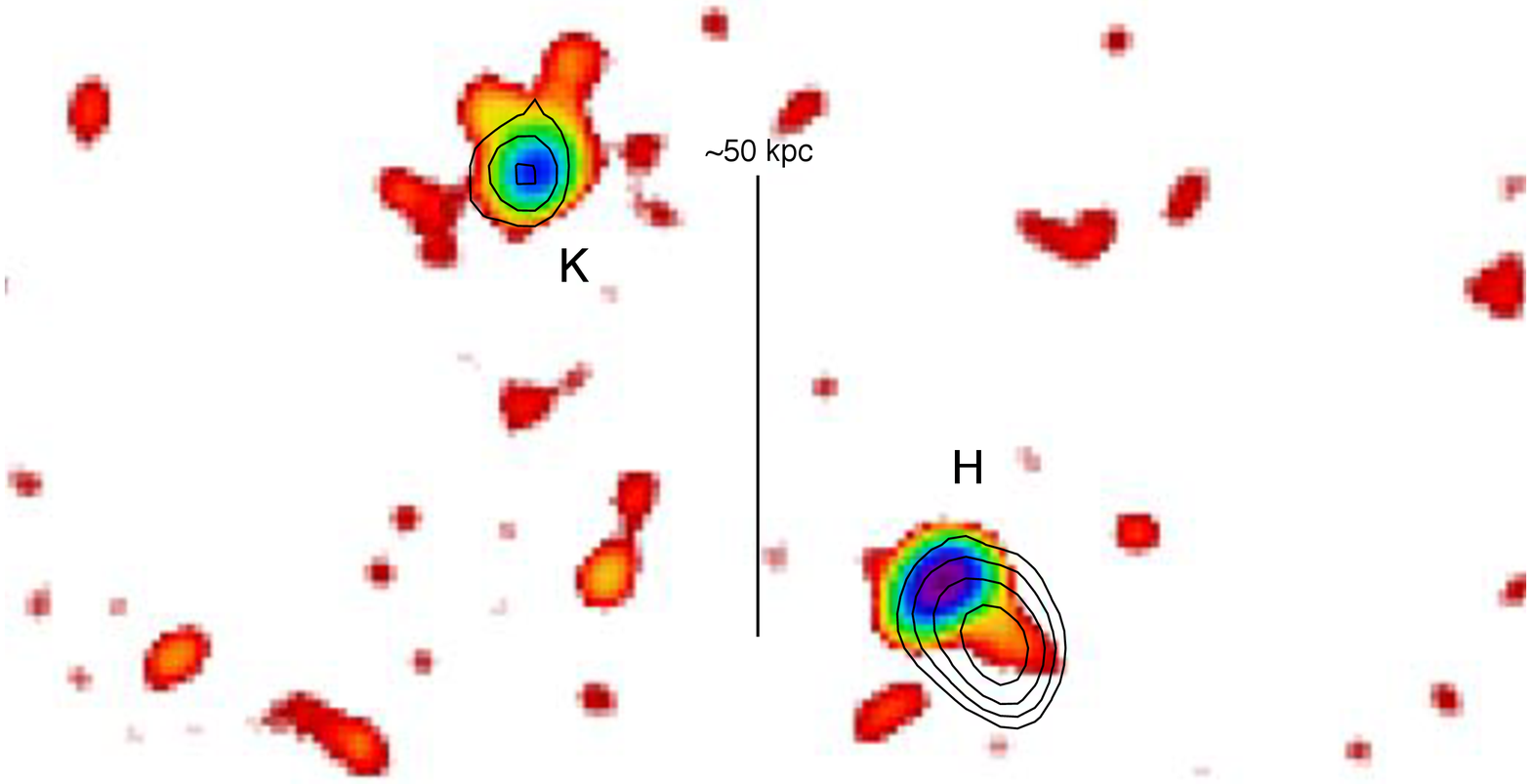}}
\caption{\small{PKS 1733$-$56, S knot and hotspot}}\label{1733_hspotS}
\vspace{1ex}
\end{subfigure}
\begin{subfigure}[b]{0.24\linewidth}
\centering
\frame{\includegraphics[trim=7.5cm 4.5cm 7cm 4cm, clip=true, angle=0.0, width=.9\linewidth]{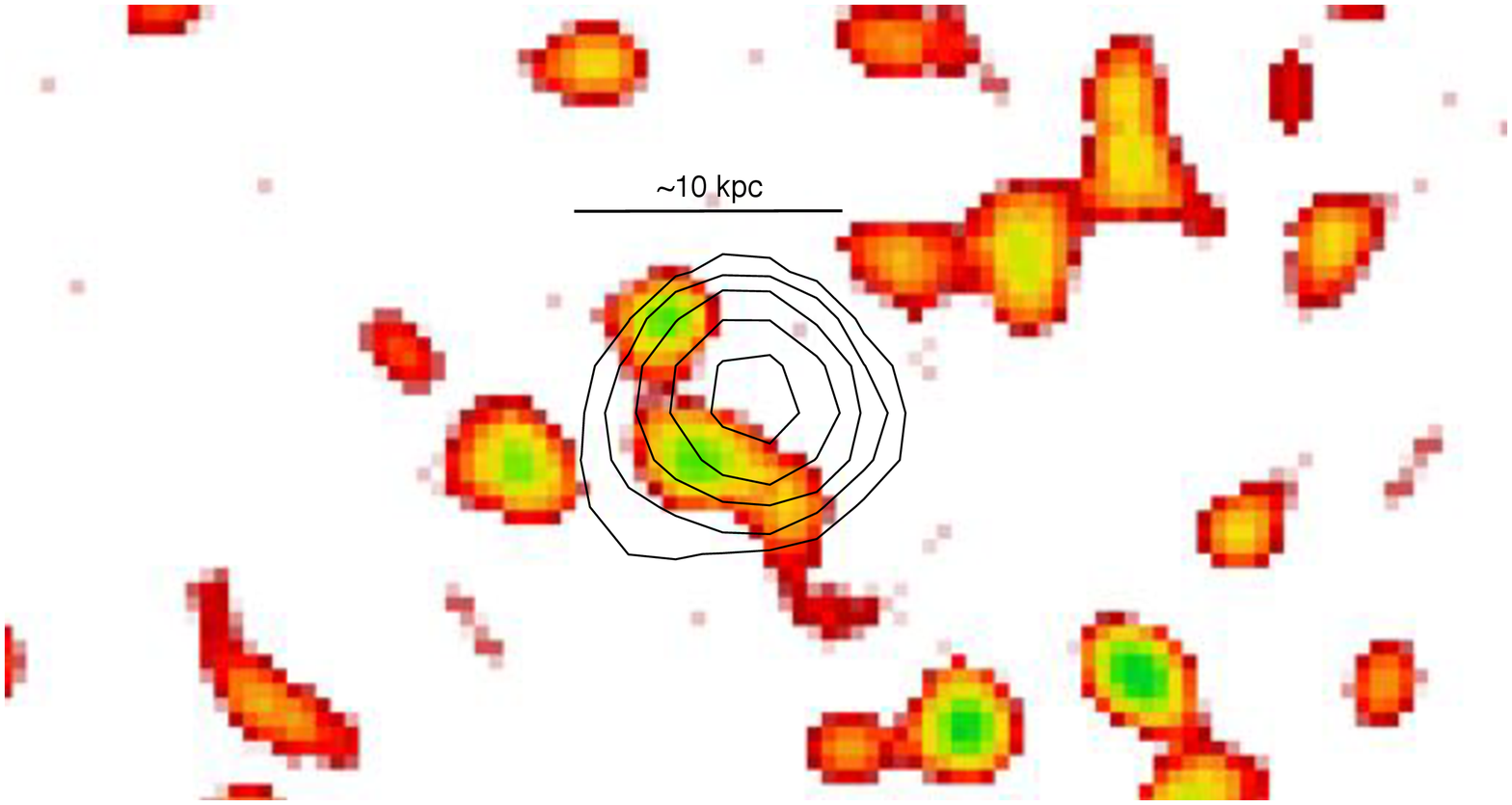}}
\caption{\small{PKS 2221$-$02, N hotspot}}\label{2221_hspotN}
\vspace{1ex}
\end{subfigure}
\begin{subfigure}[b]{0.24\linewidth}
\centering
\frame{\includegraphics[trim=7.5cm 4.5cm 7cm 4cm, clip=true, angle=0.0, width=.9\linewidth]{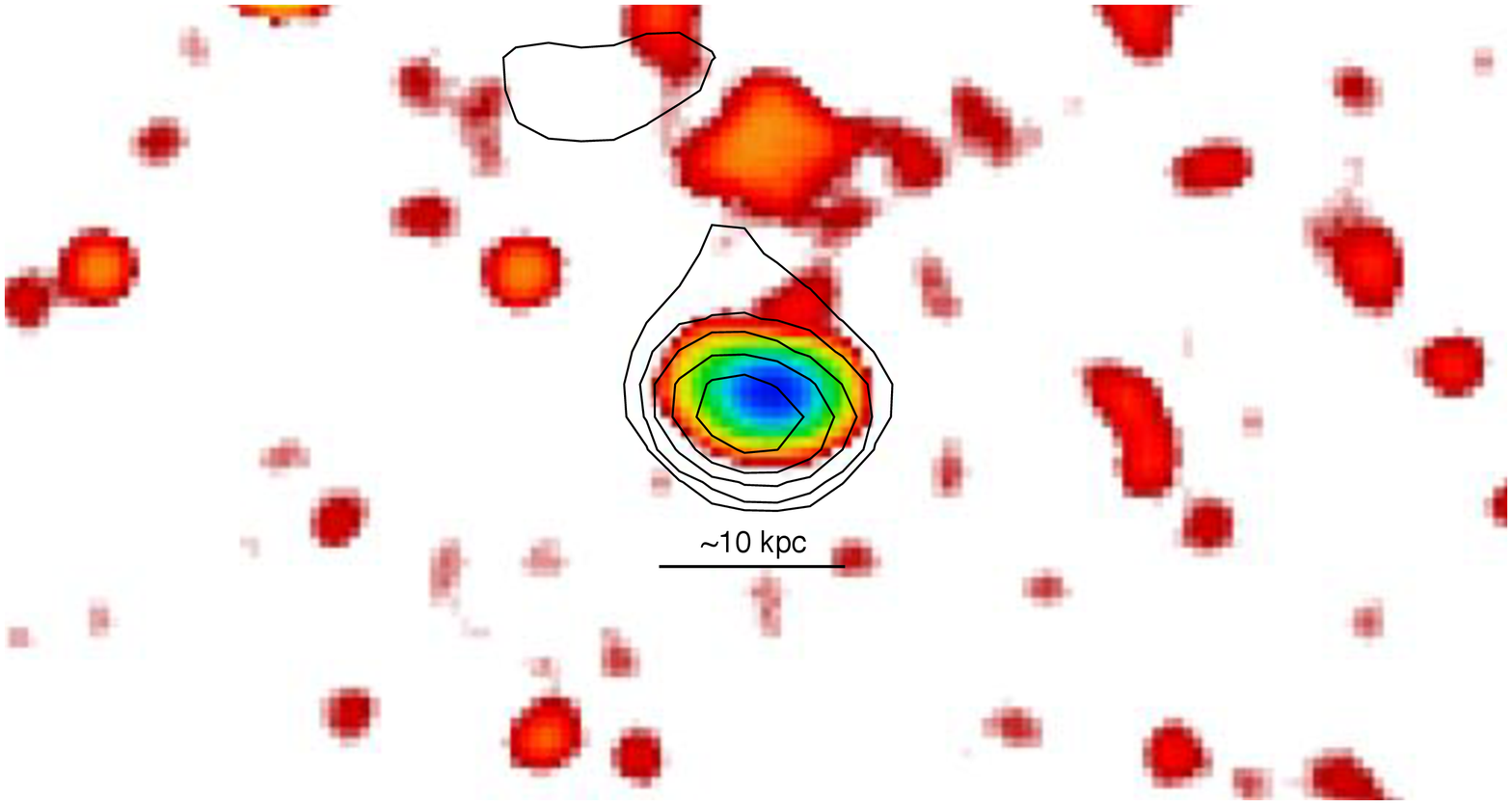}}
\caption{\small{PKS 2221$-$02, S hotspot}}\label{2221_hspotS}
\vspace{1ex}
\end{subfigure}
\begin{subfigure}[b]{0.24\linewidth}
\centering
\frame{\includegraphics[trim=7.5cm 4.5cm 7cm 4cm, clip=true, angle=0.0, width=.9\linewidth]{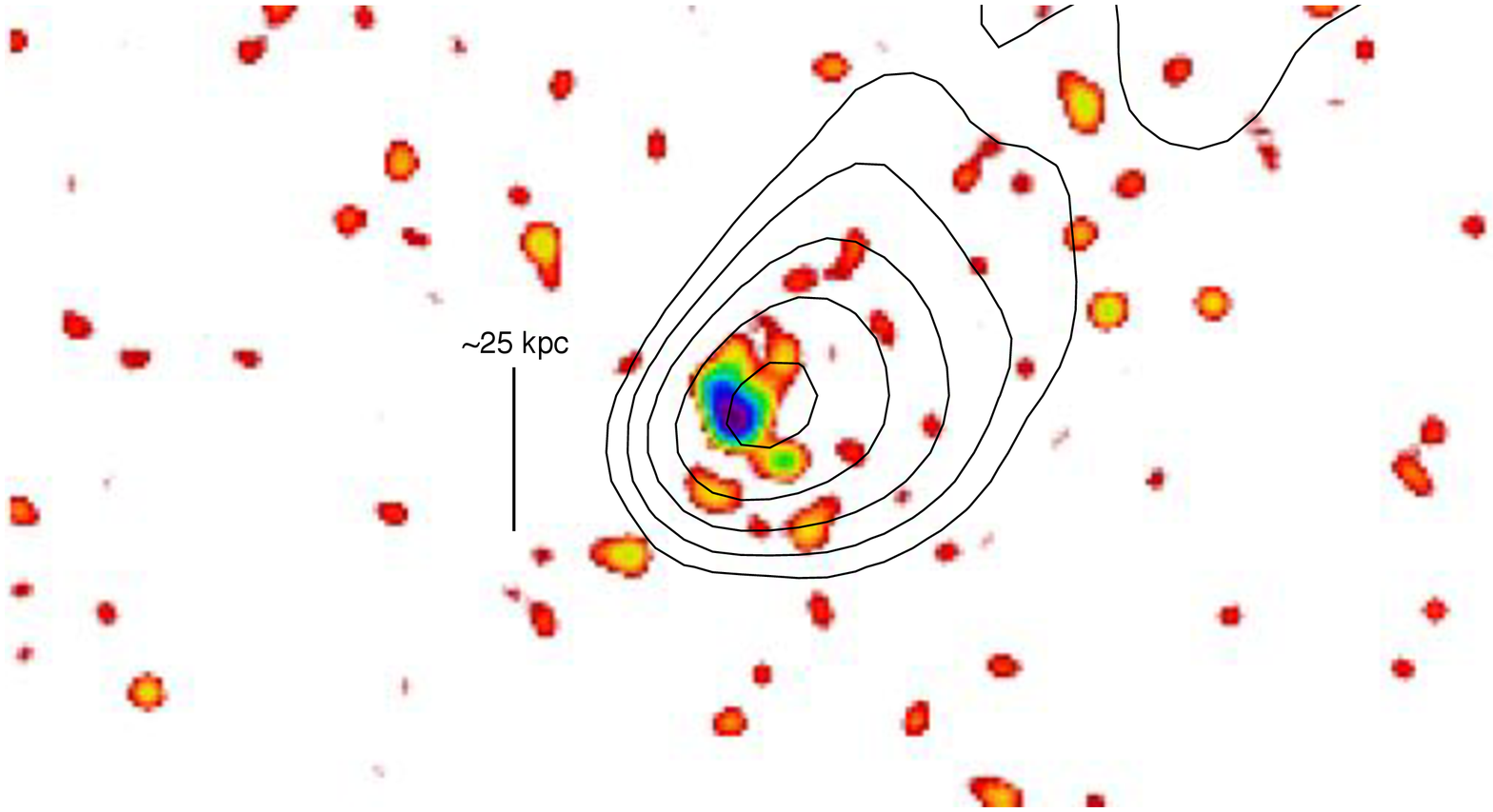}}
\caption{PKS 2356$-$61, S hotspot}\label{2356_hspotS}
\vspace{1ex}
\end{subfigure} 
\caption{\edit{X-ray hotspot close-ups for the sources in Table \ref{hotspots}}. The radio contours are the same as those detailed in the captions of Figs. \ref{2Jy_0034_f} to \ref{2Jy_2356_f}. All the images are lightly smoothed with a Gaussian profile with $\sigma=5$ pixels (1 pixel=0.492 arcsec). Where both a knot and a hotspot are present in the image, they have been labelled, respectively, as `K' and `H'. The hotspots of PKS 0945$+$07 (3C\,227), PKS 1559+02 (3C\,327), and PKS 1949+02 (3C\,403) are not displayed, as they are discussed in detail in the original publications \citep{Hardcastle2007,Kraft2005}.}\label{hotspot_fig}
\end{figure*}

We detected X-ray emission coincident with at least one of the radio hotspots and jet knots in 12 out of our 16 FRII sources (Table \ref{hotspots}), with high significance ($ \geq 3 \sigma$) in 19 out of the 23 structures listed. It has long been understood that X-ray hotpsots are very common in FRII galaxies \citep[e.g.][]{Hardcastle2007,Hardcastle2010,Massaro2010,Massaro2015}, but this is the first time that a systematic study has been carried out on a complete sample of sources. Our hotspot detection rate seems to be slightly higher than those reported in previous studies \citep[e.g.][]{Massaro2015}, but this is difficult to quantify when comparing with heterogeneous samples. Fig. \ref{hotspot_fig} shows the details of the individual detections, and it is interesting to note that in most sources there is a clear misalignment between the location of the X-ray and radio emission in at least one of the structures (knots or hotspots), on physical scales of 4--10 kpc. As mentioned in Section \ref{Introduction}, this misalignment is rather common, and it hints at complexities in the local environment or the underlying magnetic field \citep[e.g.][]{Worrall2016}.

\edit{In Table \ref{hotspots} we also tabulate the ratio between the monochromatic 1-keV X-ray flux density and the radio flux density, hereafter the X-ray/radio flux density. This quantity gives a crude characterization of the emission mechanism, with large values being more consistent with a synchrotron origin for the X-rays. We used fairly conservative regions for all the structures, allowing them to match the sizes and positions of the hotspots in the individual radio maps, adjusting them when the X-ray emission was clearly offset from the radio. We also used simple integrated fluxes, rather than background-subtracted Gaussian profile fits, as was the case for the works of \citet{Hardcastle2007} and \citet{Kraft2005}. As such, our X-ray/radio flux ratios are probably slightly smaller than those presented in the other works listed. To take the radio flux measurements we used a python plugin\footnote{\url{http://www.extragalactic.info/~mjh/radio-flux.html}} on the clean radio maps.}

The brightest newly detected X-ray hotspots in our sample are the southern hotspot and knot of PKS 1733$-$56 and the S hotspot of PKS 2356$-$61. To test whether \edit{these two hotspots} are synchrotron or inverse-Compton (synchrotron self-Compton) in origin we used the measured 1-keV flux density and the radio flux density of the corresponding hotspot to carry out inverse-Compton calculations using the code of \cite{Hardcastle1998}. As the radio maps we have are all of low resolution, we estimate the hotspot sizes \edit{for these two objects} from the fact that they appear unresolved or marginally resolved in the {\it Chandra} data, and assign all the measured radio flux density from Gaussian fitting to a spherical region of radius 1 arcsec. \edit{We use an electron energy spectrum with $\gamma_{\rm min} = 1000$ and $\gamma_{\rm max}  = 10^5$, with an energy index $p=2$ at low energies breaking to $p=3$ at $\gamma = 4000$ -- this reproduces the observed synchrotron break seen in other bright hotspots. The synchrotron spectrum is then computed between $10^4$ and $10^{12}$ Hz. The equipartition-field inverse-Compton predictions (including both SSC and inverse-Compton scattering of the CMB)} are 1.5--3 orders of magnitude below the X-ray emission observed, with the closest agreement being for PKS 2356$-$61. For this source, a field strength a factor 5 below equipartition could allow us to explain the observed X-rays as SSC emission, but this is based on the probably unrealistic assignment of 1.6 Jy of 1.4-GHz radio flux to this compact feature, and is extreme compared to other sources where SSC is the accepted explanation \citep{Hardcastle2004}. For this, and for PKS 1733$-$56 where the departure from equipartition would have to be even larger, we prefer a synchrotron model for the observed X-rays. \edit{Synchrotron models have also been applied successfully to explain the X-ray emission from hotspots in other sources, e.g. Pictor A \citep{Tingay2008}, 3C 445 \citep{Perlman2010,Orienti2012}, and 4C74.26 \citep{Erlund2010}, although the interpretation is more complicated in the latter.} 

Even considering the uncertainties, the \edit{flux density ratios for the other sources, where the lower statistics did not allow us to fit the spectra directly using monochromatic 1-keV X-ray flux density and radio flux density}, are of the same order of magnitude as those in PKS 1733$-$56 and PKS 2356$-$61, if not even larger, suggesting that an inverse-Compton emission mechanism is also unlikely in those sources. \edit{More detailed analysis would require high-resolution, multi-frequency images of the radio hotspots, which are not in general available.}

%

\section{Jets}\label{Jets}

\begin{figure}
\centering
\includegraphics[trim=6.5cm 4cm 7.2cm 4.5cm, clip=true, angle=0.0, width=0.47\textwidth]{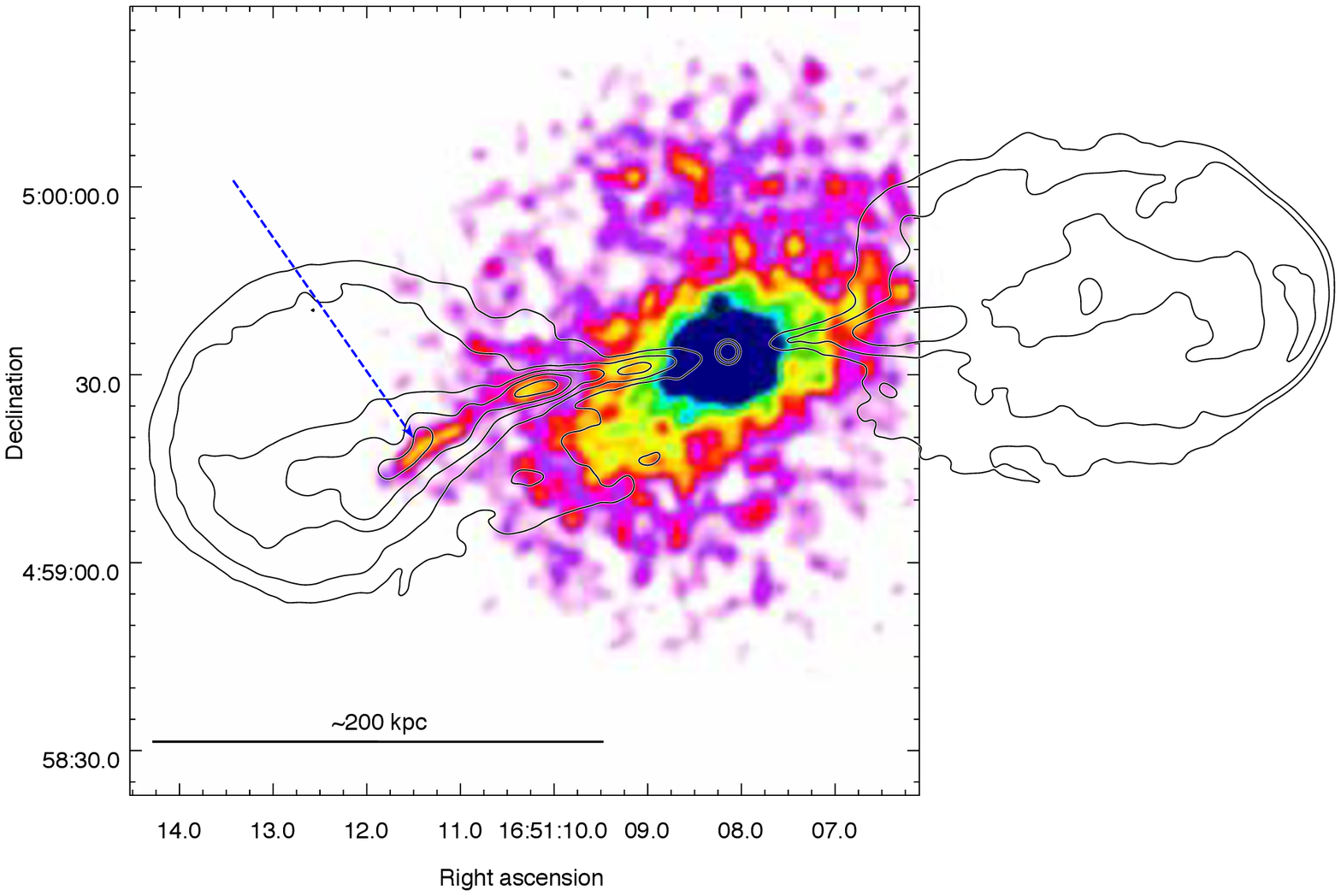}
\caption[Jet in Hercules A?]{Possible jet in PKS 1648$+$05 (3C\,348, Hercules A). Contours and radio image resolution as in Fig. \ref{2Jy_1648_f}. The blue dashed arrow indicates the enhancement in X-ray emission that might correspond to the jet.}\label{2Jy_1648_jet_f}
\end{figure}

We only detect X-ray jets clearly in two of our sources. The jet of PKS 0034$-$01 (3C\,15) is well-known, and it has been studied in detail by \citet{Dulwich2007}. Our images (Fig. \ref{2Jy_1648_jet_f}) also show evidence of a jet in PKS 1648$+$05 (3C\,348, Hercules A), extending eastwards from the nucleus, with no evidence of a counter-jet in the opposite direction. It is very possible that its existence has been noted in the past, as this source has been observed multiple times, but due to the incredibly dense and complex environment the jet is propagating through it may not have been possible to analyse it in detail.

We also observe some enhanced emission in PKS 1949$+$02 (3C\,403), Eastward from the core, which could hint at the presence of a jet, but it might arise from other mechanisms, as already pointed out by \citet{Kraft2005}.

\edit{Although it does not show clearly in our images, PKS 0521$-$36 also has an X-ray jet, which has been studied in detail by \citet{Birkinshaw2002}.}

\edit{Our results are consistent with previous studies, in terms of the number of detections of radio jets in the X-rays \citep[see e.g.][]{Sambruna2004,Jester2007,Worrall2009}. Of the 26 radio-loud AGN in our sample, 7 possess well-defined radio jets visible in our 1.4 -- 8 GHz radio maps. The X-ray jet detection fraction is therefore around 50 per cent, with 3 definite non-detections. The structure in the N lobe of PKS 0620-52 is unresolved in the radio maps, therefore it is not clear whether this source has an FRI radio jet, and although there is some excess X-ray emission in this area, it is probably linked to the dense, hot ICM. Hydra A (PKS 0915$-$11) has been extensively studied with \textit{Chandra}, but the strong X-ray ICM emission, and small angular scale of the radio jet, probably preclude its detection in the X-rays. PKS 2135-14 also shows some jet-like radio emission extending East of the core, but given the higher distance, and, comparably, lower exposure time (see Table \ref{2Jy_objects_table}), an X-ray counterpart to this structure may be too faint to be visible in our images.}

%

\section{Lobes}\label{Lobes}

\begin{table*}\small
\caption{Summary of lobe equipartition and inverse-Compton magnetic fields for the FRII sources in our sample. Further details are presented in a recent paper by \citet{Ineson2017}. In the third column, N, S, E, W, and C refer to, respectively, the north, south, east, west or both lobes, for each source (in cases where one of the lobes was split by the CCD edge, the calculations were only carried out for the lobe that was fully within the CCD). The abbreviations for the methods used to fit the X-ray data are as follows: S=fitted with free photon index ($\Gamma$); F=fitted with fixed photon index; M=modelled with unbinned data; U=upper limit. Upper/lower limits for the X-ray fluxes and inverse-Compton fields are preceded by the ``$</>$'' symbols, respectively. \edit{The 178 MHz fluxes for PKS 0404$+$03 were obtained from the data of \citet{Leahy1997}, who quote no errors; the overall flux was split into both lobes following the same ratios obtained from the 8.4 GHz data of this source.}}\label{iCompton}
\centering
\setlength{\extrarowheight}{1pt}
\begin{tabular}{cccccccccc}\hline
PKS&3C&Lobe&Radio freq.&Radio flux&Method&X-ray $\Gamma$&1 keV flux&B$_{equip.}$&B$_{inv.-Compton}$\\
&&&GHz&Jy&&&nJy&$\times10^{-10}$ T&$\times10^{-10}$ T\\\hline
0034$-$01&15&C&0.408&$9.7\pm0.3$&S&$1.9^{+0.2}_{-0.2}$&$5.0^{+0.5}_{-0.5}$&17.1&$3.89^{+0.25}_{-0.21}$\\
0038$+$09&18&C&0.408&$11.5\pm0.3$&F&1.5&$8.6^{+1.7}_{-1.7}$&12.8&$3.84^{0.52}_{0.38}$\\
0043$-$42&&N&0.408&$8.2\pm0.2$&U&1.5&$<1.0$&16.1&$>9.99$\\
&&S&0.408&$7.4\pm0.2$&U&1.5&$<0.8$&18.3&$>10.9$\\
0213$-$13&62&C&0.408&$11.7\pm0.3$&M&1.5&$4.0^{+0.7}_{-0.7}$&11.5&$5.60^{+0.62}_{-0.48}$\\
0349$-$27&&S&1.471&$1.033\pm0.002$&M&1.5&$10.9^{+1.1}_{1.1-}$&4.31&$1.05^{+0.07}_{-0.06}$\\
0404$+$03&105&N&0.178&11.130&U&1.5&$<4.7$&7.47&$>3.09$\\
&&S&0.178&8.270&U&1.5&$<4.2$&6.81&$>3.18$\\
0806$-$10&195&C&0.408&$10.2\pm0.3$&U&1.5&$<2.7$&12.3&$>6.01$\\
0945$+$07&227&W&1.429&$2.291\pm0.003$&S&$1.9^{+0.2}_{-0.2}$&$11.5^{+0.9}_{-0.9}$&7.33&$1.68^{+0.08}_{-0.07}$\\
1559$+$02&327&E&0.408&$10.8\pm0.5$&S&$1.6^{+0.5}_{-0.4}$&$6.1^{+1.4}_{-1.4}$&7.85&$3.78^{+0.60}_{-0.44}$\\
2221$-$02&445&S&1.420&$3.40\pm0.06$&S&$1.3^{+0.3}_{-0.3}$&$11.6^{+1.9}_{-1.9}$&4.52&$1.98^{+0.22}_{-0.17}$\\
2356$-$61&&N&1.472&$3.81\pm0.02$&S&$1.6^{+0.3}_{-0.2}$&$17.7^{+2.2}_{-2.2}$&7.71&$1.82^{+0.15}_{-0.12}$\\
&&S&1.472&$5.40\pm0.02$&S&$1.4^{+0.3}_{-0.2}$&$19.0^{+2.4}_{-2.4}$&8.47&$2.15^{+0.17}_{-0.14}$\\\hline
\end{tabular}
\end{table*}

We studied the lobes of the FRII sources in our sample, to find out how they compared to the results of \citet{Croston2005} in terms of their lobe pressures and equipartition (see Table \ref{iCompton}). This analysis was carried out as part of a wider FRII lobe study \citep{Ineson2017}. \edit{Full details of the method, which follows that of \citet{Croston2005}, are presented in that work, but are also summarised here. We used the radio maps to measure the radio flux densities (with the same python plugin) and determine the shapes and extent of the lobes in the X-ray images, excluding the hotspots and nuclei,} and omitting any structures that were split by the edge of the CCD (N lobe in PKS 0349$-$27, W lobe in PKS 1559$+$02, N lobe in PKS 2221$-$02), the E lobe of PKS 0945$+$07, which was contaminated by a readout streak, as well as both lobes for PKS 0442$-$28 and PKS 1733$-$56, for which the only available radio maps did not provide enough information to determine the shape and extent of the emission, both lobes of PKS 1949$+$02, which has a complex, X-shaped morphology and no apparent inverse-Compton emission, and both lobes of PKS 2211$-$17 (3C\,444), which is in a very dense and disturbed environment.

We were able to detect X-ray emission inside the lobes of eight of our sources, and to derive constraints for the rest. We assumed that the bulk of the emission originated from inverse-Compton processes, as the spectral profiles in the sources with good statistics also indicated: all the spectra were well fitted with powerlaw models (corrected for Galactic absorption), and none were improved by the addition of a thermal component, which would arise if ICM shocks were present \citep[e.g.][]{Shelton2011}. For sources with low counts, we followed the results of \citet{Croston2005} as a guideline. We then fed these results, in conjunction with the radio fluxes and lobe volumes, into the \textsc{synch} code developed by \citet{Hardcastle1998}. \textsc{synch} uses the radio spectrum and a given magnetic field to model the underlying relativistic electron population and its interaction with photons from the cosmic microwave background (CMB) and synchrotron emission. The results for an equipartition magnetic field, and one that produces the observed (inverse-Compton) X-ray emission in the lobes, are shown in Table \ref{iCompton}.

We found that all the observed magnetic fields were lower than those predicted by equipartition, although never by more than one order of magnitude. The difference in $B$ values suggests that the lobes of our FRII sources contain electron energy densities additional to the minimum energy condition, but the relatively small deviation from equipartition also suggests that our assumptions about the energetically dominant particle population in the lobes (electrons, rather than protons), are correct, all of which is consistent with the earlier results of \citet{Hardcastle2002} and \citet{Croston2005}. 

%

\section{Environments}\label{Environments}

\begin{figure*}
\centering
\begin{subfigure}[b]{0.42\linewidth}
\centering
\frame{\includegraphics[trim=5.0cm 9.0cm 6.0cm 8.0cm, clip=true, angle=-90, width=.9\linewidth]{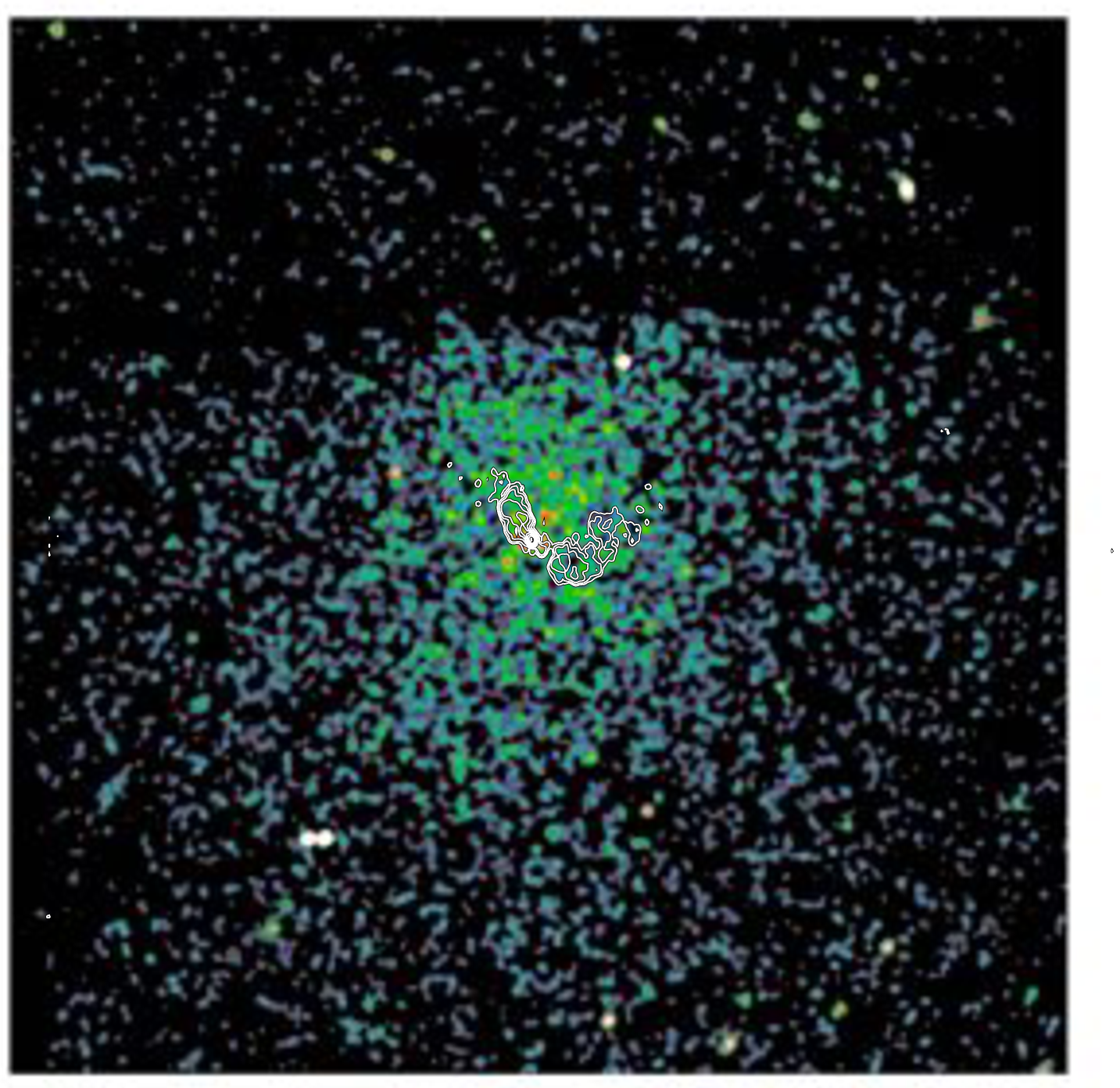}} 
\caption{PKS 0620$-$52}\label{0620_env}
\vspace{1ex}
\end{subfigure}
\begin{subfigure}[b]{0.42\linewidth}
\centering
\frame{\includegraphics[trim=5.0cm 9.0cm 6.0cm 8.0cm, clip=true, angle=-90, width=.9\linewidth]{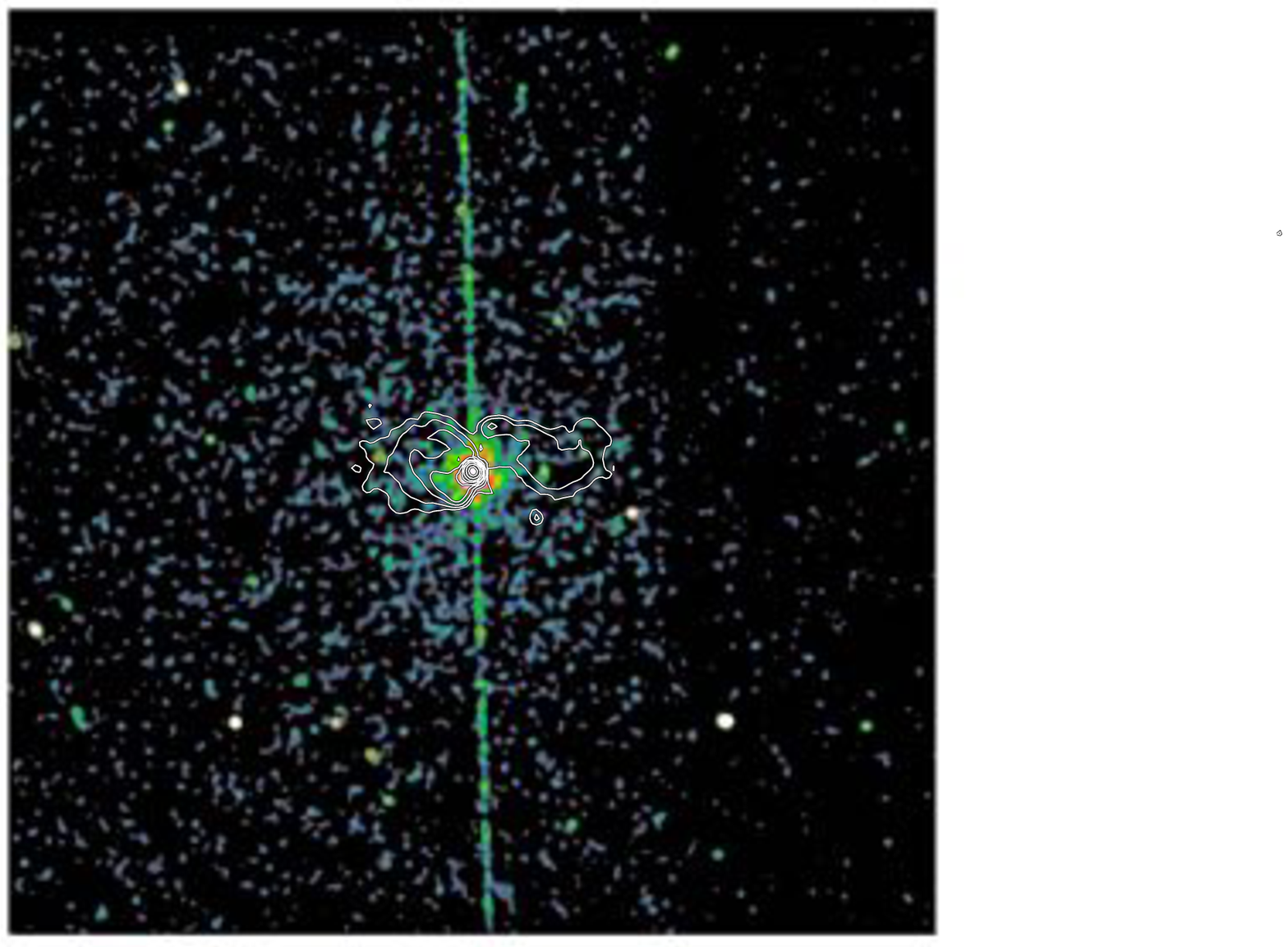}} 
\caption{PKS 0625$-$35}\label{0625_35_env}
\vspace{1ex}
\end{subfigure} 
\begin{subfigure}[b]{0.42\linewidth}
\centering
\frame{\includegraphics[trim=5.0cm 9.0cm 6.0cm 8.0cm, clip=true, angle=-90, width=.9\linewidth]{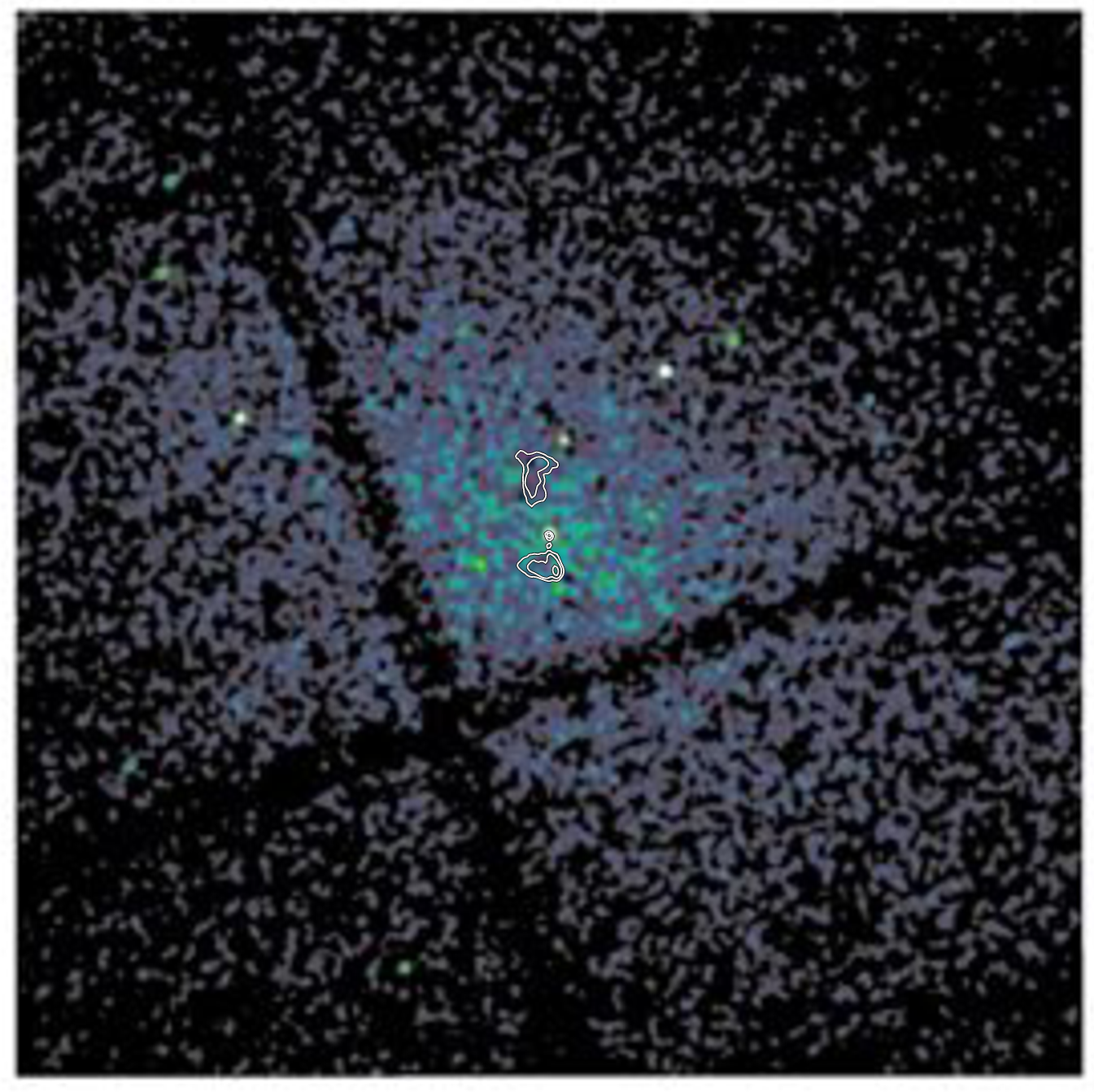}} 
\caption{PKS 0625$-$53}\label{0625_53_env}
\end{subfigure}
\begin{subfigure}[b]{0.42\linewidth}
\centering
\frame{\includegraphics[trim=5.0cm 9.0cm 6.0cm 8.0cm, clip=true, angle=-90, width=.9\linewidth]{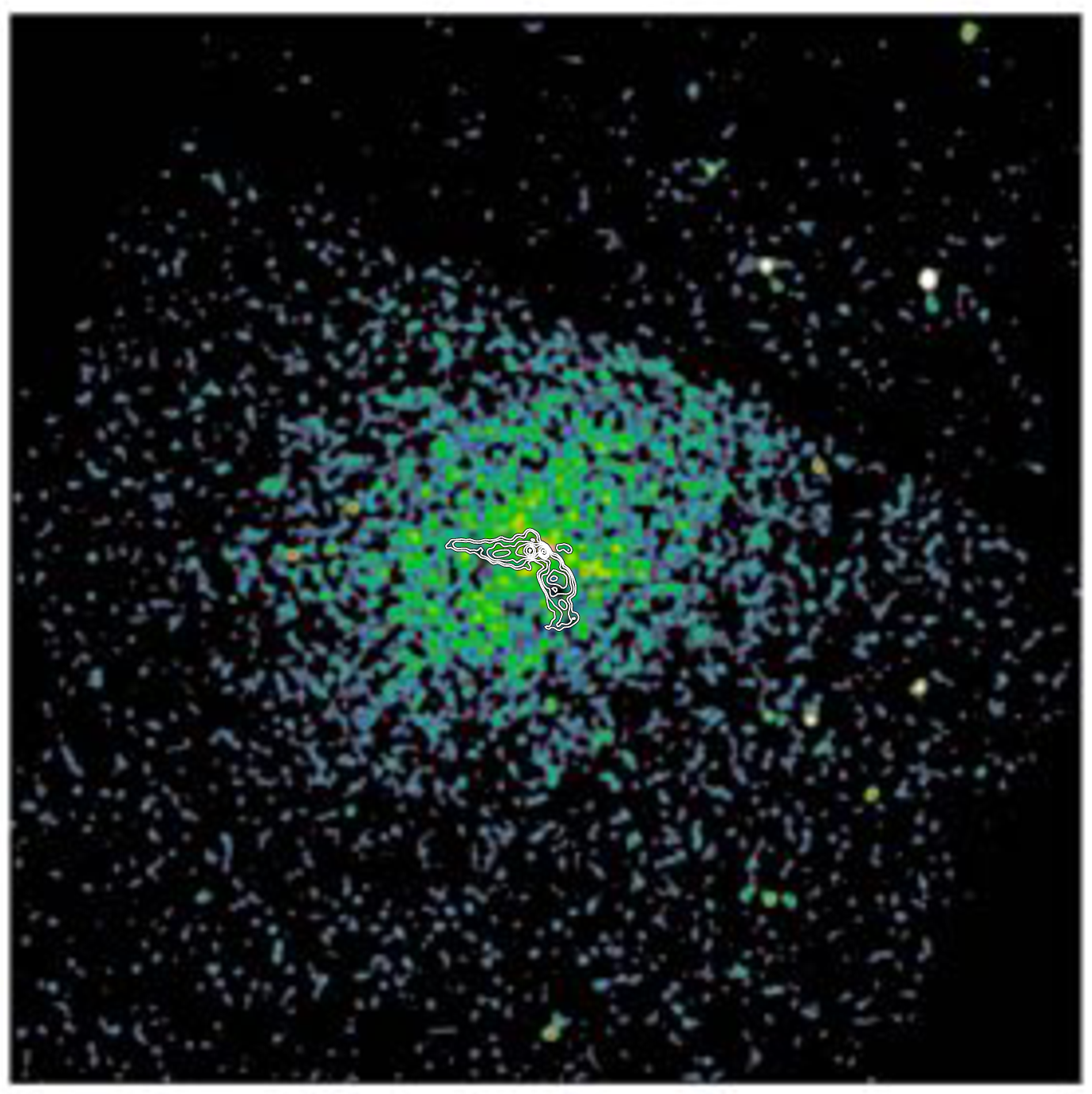}} 
\caption{PKS 1839$-$48}\label{1839_env}
\end{subfigure}
\begin{subfigure}[b]{0.42\linewidth}
\centering
\frame{\includegraphics[trim=5.0cm 9.0cm 6.0cm 8.0cm, clip=true, angle=-90, width=.9\linewidth]{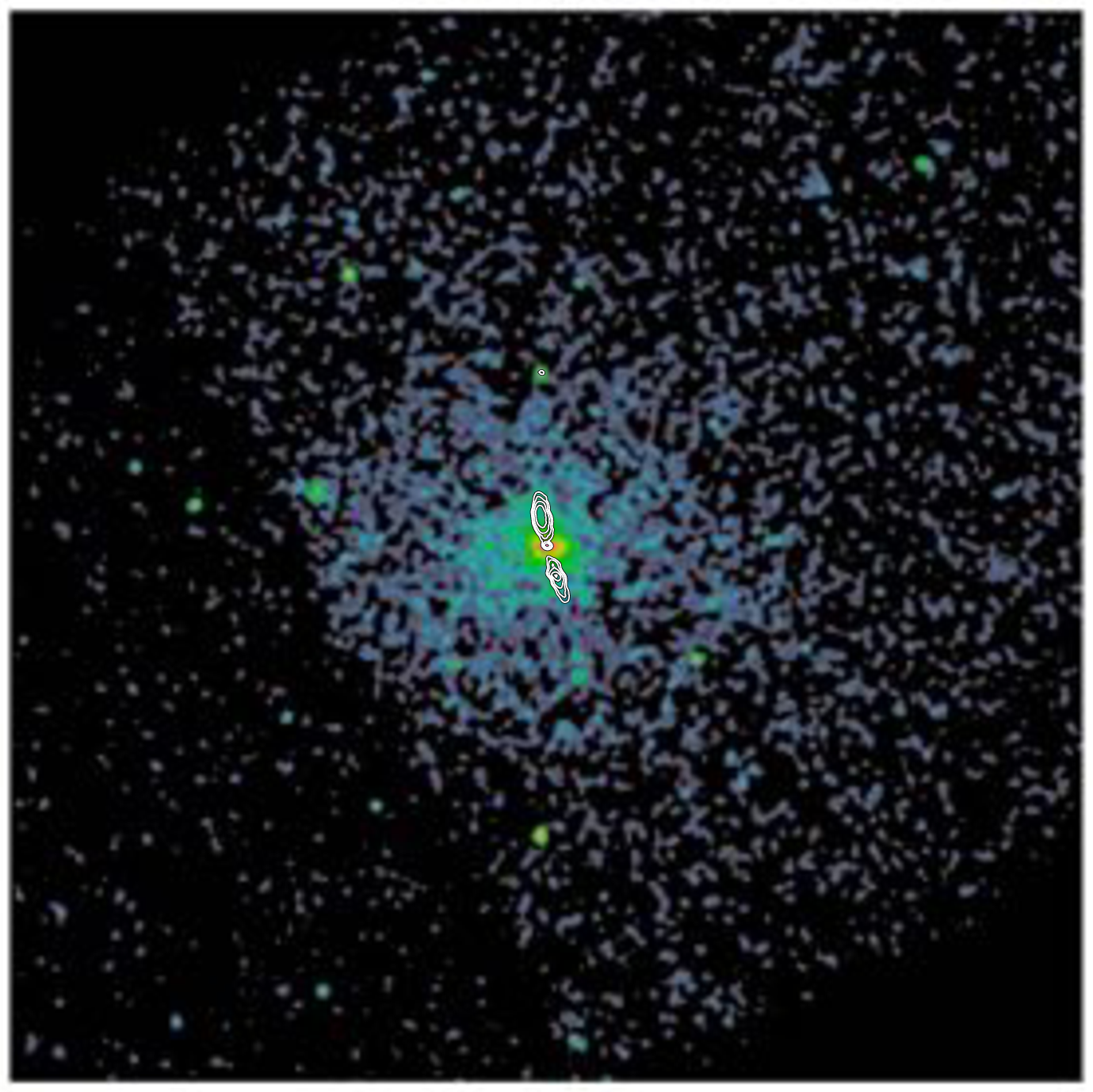}} 
\caption{PKS 1954$-$55}\label{1954_env}
\end{subfigure}
\caption{Zoomed-out X-ray images illustrating the dense environments of the FRI LERGs in the low-$z$ 2Jy sample. All the images are smoothed with a Gaussian profile with $\sigma=7$ pixels (1 pixel=0.492 arcsec). PKS 0915$-$11 (Hydra A) and PKS 1648$+$05 (Hercules A) are not included, as their environments can already be clearly seen in Figs. \ref{2Jy_0915_f} and \ref{2Jy_1648_f}. We have not included here the FRII LERGs, as the environment of PKS 2211$-$17 (3C\,444) can be clearly seen in Fig. \ref{2Jy_2211_f}, and PKS 0034$-$01 (3C\,15) and PKS 0043$-$42, whose nuclear spectra are atypical for LERGs, show no clear signs of extended emission.}\label{environments_f} 
\end{figure*}

\citet{Ineson2013,Ineson2015} found that the environments of radio-loud AGN are different depending on their accretion mode. They found that for LERGs, most of which are FRI, there is a correlation between radio luminosity and ICM richness, while no correlation was apparent for HERGs (high excitation radio galaxies, the radiatively efficient sources), and they seemed to avoid the richest environments. All seven of our FRI sources are LERGs, in the upper range of the FRI radio power distribution. They all show clear evidence of large-scale extended X-ray emission around the host (with PKS 0625$-$35 having the poorest environment among them, see Fig. \ref{environments_f}), and several of them inhabit well-known clusters. We would expect lower luminosity LERGs to be found in poorer environments, but they aren't represented in the 2Jy sample.

Of our 16 FRII sources, three are classified as LERGs: PKS 0034$-$01 (3C\,15), PKS 0043$-$42, and PKS 2211$-$17 (3C\,444). The first two sources, however, have X-ray spectra that are somewhat atypical for LERGs, and PKS 0043$-$42, in particular, shows signs of radiatively efficient accretion (\citealt{RamosAlmeida2011b}; \citetalias{Mingo2014}). PKS  2211$-$17 is a bona-fide LERG, and it inhabits a well-known cluster. PKS 0043$-$42 shows signs of extended X-ray emission, which \citet{Ineson2015} found to be consistent with a weak cluster or group environment. There are no signs of extended emission around PKS 0034$-$01, and \citet{RamosAlmeida2013b} found only a weak environment around it.

Of the 13 HERG FRII, only three (PKS 0349$-$27, PKS 1733$-56$, and PKS 1949$+$02) show some traces of extended X-ray emission \citep[see also][]{Ineson2015}. However, several of the HERG FRII sources present some smaller-scale, low surface brightness extended emission around the nucleus or the edges of the lobes, and in the optical, far from being isolated, many of them have dense environments, close companions, or show signs of recent interaction \citep{RamosAlmeida2011,RamosAlmeida2013b}. It is possible that we are not detecting their extended ICM emission in the X-rays because the HERGs in our sample are found, on average, at higher $z$ than the LERGs.

An extended, quantitative analysis of the 2Jy environments has been presented by \citet{Ineson2015}, as part of their broader study of the properties of radio galaxies. \citet{Ineson2017} also present a detailed analysis of the pressure balance between the FRII sources in our sample and their environments, in the context of a larger FRII sample. Here we just note that the Mach numbers for the expansion of the lobes of the 2Jy sources, obtained by considering the Rankine-Hugoniot conditions at the lobe tip, are found in their analysis to be in the range 1 to 3, with an average Mach number $\sim2.1$. This is similar to the Mach numbers of comparable systems \citep[e.g.]{Croston2011,Shelton2011,Kraft2012,Harwood2016}, but lower than those we obtained for lower-power systems in less dense environments \citep[e.g.][]{Kraft2003,Croston2007,Mingo2011,Mingo2012}, which is expected.

%

\section{Conclusions}\label{Conclusions}

In agreement with previous results, we find that X-ray hotspots and jet knots are fairly ubiquitous in FRII galaxies, with at least one of them being detected in 12 out of our 16 sources, with high significance ($\sigma \geq 3$) in all but four out of the 23 structures detected (listed in Table \ref{hotspots}). We also observe a clear misalignment between the radio and X-ray emission in several sources, on physical scales of 4--10 kpc.

The hotspots whose spectra we have been able to fit show, invariably, synchrotron emission spectra. Our calculations for PKS 1733$-$56 and PKS 2356$-$61 show that inverse-Compton emission is unlikely.

We only observed jets unequivocally in two of our sources, PKS 0034$-$01 (3C\,15), and PKS 1648$+$05 (3C\,348, Hercules A). 

We found that the lobes of all the FRII sources in our sample have magnetic fields that are lower than expected from equipartition conditions, though never by more than an order of magnitude. These results are consistent with those of previous studies of similar sources.

We also confirmed the tendency of luminous LERGs (mostly FRI) to inhabit rather dense environments, consistent with the results of \citet{RamosAlmeida2013b} and \citet{Ineson2015}, while our HERGs (mostly FRII) seem to inhabit slightly sparser areas.

\subsection*{Acknowledgements}

\edit{We thank the anonymous referee for their constructive comments, which have improved the paper}. We thank J. P. Leahy for providing the radio map for Hercules A (PKS 1648$+$05). BM acknowledges support from the UK Space Agency. MJH acknowledges support from the Science \& Technology Facilities Council (STFC; grant number ST/M001008/1). RM gratefully acknowledges support from the European Research Council under the European Union's Seventh Framework Programme (FP/2007-2013) /ERC Advanced Grant RADIOLIFE-320745. This work has made use of new and archival data from \textit{Chandra} and software provided by the Chandra X-ray Center (CXC) in the application package CIAO. This work also makes use of data from The Australia Telescope Compact Array (ATCA), which is part of the Australia Telescope National Facility, funded by the Australian Government for operation as a National Facility managed by CSIRO, as well as data from the Karl G. Jansky Very Large Array (VLA), part of the The National Radio Astronomy Observatory, a facility of the National Science Foundation operated under cooperative agreement by Associated Universities, Inc.

\bibliographystyle{mnras}
\bibliography{2Jy_2}

\end{document}